# The reconstructed thermal lattice Boltzmann flux solver and its applications for simulations of thermal flows

Jinhua Lu[1], Chuanshan Dai[3,4], Peng Yu[1,2,5,*]

[1] Department of Mechanics and Aerospace Engineering, Southern University of Science and Technology, Shenzhen 518055, China.

[2] Guangdong Provincial Key Laboratory of Turbulence Research and Applications, Southern University of Science and Technology, Shenzhen 518055, China.

[3] Key Laboratory of Ministry of Education on Efficient Utilization of Low and Medium Grade Energy, Tianjin University, Tianjin, 300350, China

[4] School of Mechanical Engineering, Tianjin University, Tianjin, 300350, China,

[5] Center for Complex Flows and Soft Matter Research, Southern University of Science and Technology, Shenzhen, 518055, China

## Abstract

The thermal lattice Boltzmann flux solver (TLBFS) has been proposed to overcome the drawbacks of the thermal lattice Boltzmann models. However, as a weakly compressible model, its mechanism of good numerical stability for high Rayleigh number thermal flows is still unclear. To reveal the mechanism, the present paper firstly derives the macroscopic equations of TLBFS (MEs-TLBFS) with actual numerical dissipation terms by approximating its computational process. By solving MEs-TLBFS with the finite volume method, the reconstructed TLBFS (RTLBFS) is proposed. Detailed analyses prove that these actual numerical dissipation terms are the mechanism of the good numerical stability of TLBFS for high Rayleigh number thermal flows. More detailed numerical tests indicate RTLBFS has similar performances as TLBFS for stability, accuracy, and efficiency. Moreover, the present RTLBFS shows a clear mechanism to achieve good numerical stability for high Rayleigh number flows.

## 1. Introduction

The lattice Boltzmann equation is a mesoscopic model that provides a general description for non-equilibrium physical processes, such as fluid flow and heat/mass transfer. Due to the simple implementation, the lattice Boltzmann model has received much attention. A widely-used lattice Boltzmann model, i.e., the D$m$Q$n$ model [1], was proposed to simulate incompressible flow. Subsequently, a series of thermal lattice Boltzmann models have also been proposed to simulate incompressible thermal flows. The current thermal lattice Boltzmann models can be roughly classified into three categories, i.e., the multi-speed model, the double distribution function (DDF)

Corresponding author, E-mail: P. Yu (yup6@sustech.edu.cn)

model, and the hybrid lattice Boltzmann model.

By introducing additional discrete lattice speeds, the multi-speed model [2-5] uses one distribution function to describe both the flow and energy evolutions. However, the Prandtl number cannot be modified in the earliest multi-speed models [2, 3], which is inconsistent with the real fluid. This limitation has been overcome by some improved multi-speed models [4, 5] with adjustable Prandtl numbers. In multi-speed models, the additional discrete lattice speeds lead to complicated equilibrium distribution functions and pose extra difficulties in the implementations of boundary conditions.

The DDF model adopts two distribution functions to separately describe the flow and energy evolutions. In the complete DDF models, compression work and viscous heat dissipation can be taken into account [6, 7]. Since the compression work and viscous heat dissipation can be neglected for incompressible laminar flows in general, some simplified models [8-11] that neglect compression work and viscous heat dissipation have been proposed as well.

The hybrid lattice Boltzmann model [12-15] adopts the standard lattice Boltzmann equation [1] to simulate the flow field and adopts the macroscopic energy equation to simulate the temperature field. In general, the macroscopic energy equation in these models is solved by the finite difference method.

All the thermal lattice Boltzmann models mentioned above apply the lattice Boltzmann method (LBM) to simulate thermal fluid flows, which thus inherit the same drawbacks in LBM. Firstly, these models have to be numerically solved on uniform mesh in general. Complicated and time-consuming interpolation calculations are required if the non-uniform mesh is adopted [16, 17]. Another drawback is the coupled time step with mesh size. Thus, the implementation of the adaptive and multi-block computations becomes complicated [18, 19]. The third drawback is the extra memory size for distribution functions. Except for the macroscopic variables, distribution functions need to be stored as well. It may be a heavy burden for three-dimensional cases with a large amount of mesh.

A few attempts have been made to overcome these drawbacks of thermal lattice Boltzmann models. Chen et al. [20-23] proposed a simplified thermal LBM (STLBM) [20-23], which only involves macroscopic variables. Therefore, STLBM needs less memory size, and the boundary conditions are easy to implement. STLBM can be adapted to the non-uniform mesh by using the Lagrange interpolation [23]. Consequently, the limitations of uniform mesh and coupled time step with mesh size can be overcome as well. Wang et al. [24, 25] developed a thermal lattice Boltzmann flux solver (TLBFS), which was derived from a DDF model by rewriting the expanded equations in the finite volume framework. The interface fluxes can be obtained by reconstructing the local solution of the thermal lattice Boltzmann model. Since TLBFS is a finite volume solver, the limitation of uniform mesh and the coupled time step with mesh size can be easily overcome. Besides, TLBFS only involves equilibrium distribution functions expressed by macroscopic variables, and thus only macroscopic variables need to be stored. TLBFS has been proven to be both accurate and efficient, and has good numerical stability for high Rayleigh number flows [26].

TLBFS is a promising approach that overcomes the drawbacks of thermal lattice Boltzmann models, and the mathematical analysis [24] indicates that TLBFS can recover the macroscopic continuity, Navier-Stokes, and energy equations with second-order accuracy. Since TLBFS only involves macroscopic variables, it can be considered a macroscopic weakly compressible model. However, the previous study [27] indicates that directly solving weakly compressible models based on the central difference scheme, in general, suffers serious numerical stability. Additional treatments, such as the preconditioned method [28, 29] or adding some numerical dissipative terms [30, 31], are needed to stabilize numerical computation. In contrast, TLBFS shows good numerical stability for high Rayleigh number flows, but the associated mechanism is still unclear. Therefore, revealing the mechanism in the perspective of macroscopic scale motivates the present work.

Recently, Lu et al. [27] proposed an idea to recover the time-discretized macroscopic equations (MEs) from the lattice Boltzmann equation by approximating the actual computational process. Inspired by the idea, in the present paper, the MEs-TLBFS with actual dissipation terms are derived firstly by approximating its actual computational process. By solving MEs-TLBFS with the finite volume method (FVM), RTLBFS is proposed. The mechanism of the good numerical stability of TLBFS for high Rayleigh number flows is revealed by analyzing RTLBFS. Detailed numerical investigations prove that RTLBFS can recover the numerical stability and accuracy of TLBFS very well, and can be flexibly applied on non-uniform meshes with curved boundaries as well.

The remaining parts of this paper are organized as follows. Section 2 introduces TLBFS. Section 3 proposes RTLBFS and analyzes the mechanisms of the good numerical stability of TLBFS. In Section 4, five benchmark tests are simulated to validate RTLBFS. Finally, some conclusions are summarized in Section 5.

## 2. Derivation of MEs-TLBFS

### 2.1 Thermal lattice Boltzmann model

The derivation of TLBFS is based on the DDF thermal lattice Boltzmann model without compression work and viscous heat dissipation [9], which can be written as

$$f_i(\mathbf{x}+\mathbf{e}_i\delta t, t+\delta t) = f_i(\mathbf{x},t) - \frac{1}{\tau_f}\left[f_i(\mathbf{x},t) - f_i^{eq}(\mathbf{x},t)\right], \tag{1}$$

$$g_i(\mathbf{x}+\mathbf{e}_i\delta t, t+\delta t) = g_i(\mathbf{x},t) - \frac{1}{\tau_g}\left[g_i(\mathbf{x},t) - g_i^{eq}(\mathbf{x},t)\right], \tag{2}$$

where $\mathbf{x}$ is the position of the computational node, $\mathbf{e}_i$ are the discrete velocities; $t$ is the time; $\delta t$ is the time interval; $f_i$ and $f_i^{eq}$ are the density distribution functions and equilibrium density distribution functions, respectively; $g_i$ and $g_i^{eq}$ are the energy distribution functions and equilibrium energy distribution functions, respectively; $\tau_f$ and $\tau_g$ are the relaxation parameters for $f_i$ and $g_i$, respectively; $\delta t$ is the time interval. The equilibrium distribution functions $f_i^{eq}$ and $g_i^{eq}$ are

$$f_i^{eq} = w_i\rho\left[1 + \frac{\mathbf{e}_i\cdot\mathbf{u}}{c_s^2} + \frac{(\mathbf{e}_i\cdot\mathbf{u})^2}{2c_s^4} - \frac{\mathbf{u}^2}{2c_s^2}\right], \tag{3}$$

$$g_i^{eq} = w_i\rho c_p T\left(1 + \frac{\mathbf{e}_i\cdot\mathbf{u}}{c_s^2}\right), \tag{4}$$

where $w_i$, $\rho$, $c_p$, $\mathbf{u}$, $c_s$ and $T$ are the weight coefficient, density, specific heat capacity at constant pressure, velocity, sound speed, and temperature, respectively. For the 2D and 3D situations, the D2Q9 model and the D3Q19 model are adopted respectively. The discrete velocities of the D2Q9 and D3Q19 models are

$$\text{D2Q9}: \mathbf{e}_i = \begin{bmatrix} 0 & 1 & 0 & -1 & 0 & 1 & -1 & -1 & 1 \\ 0 & 0 & 1 & 0 & -1 & 1 & 1 & -1 & -1 \end{bmatrix} c,$$

$$\text{D3Q19}: \mathbf{e}_i = \begin{bmatrix} 0 & 1 & -1 & 0 & 0 & 0 & 0 & 1 & -1 & 1 & -1 & 1 & -1 & 1 & -1 & 0 & 0 & 0 & 0 \\ 0 & 0 & 0 & 1 & -1 & 0 & 0 & 1 & 1 & -1 & -1 & 0 & 0 & 0 & 0 & 1 & -1 & 1 & -1 \\ 0 & 0 & 0 & 0 & 0 & 1 & -1 & 0 & 0 & 0 & 0 & 1 & 1 & -1 & 1 & 1 & 1 & -1 & -1 \end{bmatrix} c$$

(5)

The corresponding $w_i$ and $c_s$ are

$$\begin{cases} \text{D2Q9}: & w_0 = 4/9, \quad w_{1-4} = 1/9, \quad w_{5-8} = 1/36 \quad c_s = c/\sqrt{3} \\ \text{D3Q19}: & w_0 = 1/3, \quad w_{1-6} = 1/18, \quad w_{7-18} = 1/36 \quad c_s = c/\sqrt{3} \end{cases}. \quad (6)$$

The equilibrium distribution functions satisfy

$$\sum_i f_i^{eq} = \rho, \quad \sum_i f_i^{eq} e_{i\alpha} = \rho u_\alpha, \quad \sum_i f_i^{eq} e_{i\alpha} e_{i\beta} = \rho c_s^2 \delta_{\alpha\beta} + \rho u_\alpha u_\beta,$$

$$\sum_i f_i^{eq} e_{i\alpha} e_{i\beta} e_{i\gamma} = \rho c_s^2 \left( u_\alpha \delta_{\beta\gamma} + u_\beta \delta_{\alpha\gamma} + u_\gamma \delta_{\alpha\beta} \right), \quad (7)$$

$$\sum_i g_i^{eq} = \rho c_p T, \quad \sum_i g_i^{eq} e_{i\alpha} = \rho c_p u_\alpha T, \quad \sum_i g_i^{eq} e_{i\alpha} e_{i\beta} = \rho c_p T c_s^2 \delta_{\alpha\beta}. \quad (8)$$

The relaxation parameters $\tau_f$ and $\tau_g$ are respectively related to the kinematic viscosity $\upsilon$ and thermal diffusivity coefficient $\chi$ through

$$\upsilon = (\tau_f - 0.5) c_s^2 \delta t, \quad (9)$$

$$\chi = (\tau_g - 0.5) c_s^2 \delta t. \quad (10)$$

The macroscopic variables $\rho$, $u_\alpha$ and $T$ are respectively recovered by

$$\rho = \sum_i f_i, \quad (11)$$

$$\rho u_\alpha = \sum_i f_i e_{i\alpha}, \quad (12)$$

$$\rho c_p T = \sum_i g_i. \quad (13)$$

The pressure is determined by the equation of state

$$p = \rho c_s^2. \quad (14)$$

## 2.2 TLBFS

The derivation of TLBFS is based on the Chapman-Enskog expansion analysis. By using the second-order Taylor expansion, Eqs. (1) and (2) can be written in a continuous form as

$$(\partial_t + e_{i\alpha} \partial_\alpha) f_i + \frac{\delta t}{2} (\partial_t + e_{i\alpha} \partial_\alpha)^2 f_i + O(\delta t^2) = -\frac{1}{\tau_f \delta t} (f_i - f_i^{eq}), \quad (15a)$$

$$(\partial_t + e_{i\alpha} \partial_\alpha) g_i + \frac{\delta t}{2} (\partial_t + e_{i\alpha} \partial_\alpha)^2 g_i + O(\delta t^2) = -\frac{1}{\tau_g \delta t} (g_i - g_i^{eq}). \quad (15b)$$

Then, $f_i$ and $g_i$ are respectively expanded as

$$f_i = f_i^{(0)} + \varepsilon f_i^{(1)} + \varepsilon^2 f_i^{(2)} + O(\varepsilon^3), \tag{16a}$$

$$g_i = g_i^{(0)} + \varepsilon g_i^{(1)} + \varepsilon^2 g_i^{(2)} + O(\varepsilon^3), \tag{16b}$$

where $\varepsilon$ is a small parameter. Simultaneously, two time scales $t1 = \varepsilon t$, $t2 = \varepsilon^2 t$ and a length scale $x1 = \varepsilon x$ are introduced, and thus we have

$$\partial_t = \varepsilon \partial_{t1} + \varepsilon^2 \partial_{t2}, \quad \partial_\alpha = \varepsilon \partial_{1\alpha}. \tag{17}$$

Substituting Eqs. (16a), (16b) and (17) into Eqs. (15a) and (15b), Eqs. (15a) and (15b) can be rewritten in the different orders of $\varepsilon$ as follows:

$$\varepsilon^0 : f_i^{(0)} = f_i^{eq}, \tag{18a}$$

$$\varepsilon^0 : g_i^{(0)} = g_i^{eq}, \tag{18b}$$

$$\varepsilon^1 : (\partial_{t1} + e_{i\alpha}\partial_{1\alpha}) f_i^{(0)} = -\frac{1}{\tau_f \delta t} f_i^{(1)}, \tag{19a}$$

$$\varepsilon^1 : (\partial_{t1} + e_{i\alpha}\partial_{1\alpha}) g_i^{(0)} = -\frac{1}{\tau_g \delta t} g_i^{(1)}, \tag{19b}$$

$$\varepsilon^2 : \partial_{t2} f_i^{(0)} + \left(1 - \frac{1}{2\tau_f}\right)(\partial_{t1} + e_{i\alpha}\partial_{1\alpha}) f_i^{(1)} = -\frac{1}{\tau_f \delta t} f_i^{(2)}, \tag{20a}$$

$$\varepsilon^2 : \partial_{t2} g_i^{(0)} + \left(1 - \frac{1}{2\tau_g}\right)(\partial_{t1} + e_{i\alpha}\partial_{1\alpha}) g_i^{(1)} = -\frac{1}{\tau_g \delta t} g_i^{(2)}. \tag{20b}$$

Combining Eqs. (7), (11), (12), (16a), and (18a) leads to

$$\varepsilon^n : \sum_i f_i^{(n)} = 0, \quad \sum_i f_i^{(n)} e_{i\alpha} = 0, \quad n \geq 1. \tag{21}$$

Similarly, combining Eqs. (8), (13), (16b), and (18b) leads to

$$\varepsilon^n : \sum_i g_i^{(n)} = 0, \quad n \geq 1. \tag{22}$$

Taking the summations of Eqs. (19a) and (20a) over $i$ yields

$$\partial_t \rho + \partial_\alpha \left( \sum_i e_{i\alpha} f_i^{eq} \right) = 0. \tag{23}$$

Multiplying both sides of Eqs. (19a) and (20a) by $e_{i\beta}$ and taking the summations over $i$ leads to

$$\partial_t (\rho u_\alpha) + \partial_\beta \left\{ \sum_i \left[ e_{i\alpha} e_{i\beta} f_i^{eq} + \left(1 - \frac{1}{2\tau_f}\right) e_{i\alpha} e_{i\beta} f_i^{(1)} \right] \right\} = 0 \tag{24}$$

Taking the summations of Eqs. (19b) and (20b) over $i$ yields

$$\partial_t (\rho c_p T) + \partial_\alpha \left\{ \sum_i \left[ e_{i\alpha} g_i^{eq} + \left(1 - \frac{1}{2\tau_g}\right) e_{i\alpha} g_i^{(1)} \right] \right\} = 0 \tag{25}$$

According to the Gauss theorem, Eqs. (23), (24), and (25) can also be written in a spatially discretized scheme as

$$\partial_t \rho + \frac{1}{\Delta V} \sum_k \left( \sum_i e_{i\alpha} f_i^{eq} \right) \Delta S_{k\alpha} = 0 \tag{26}$$

$$\partial_t (\rho u_\alpha) + \frac{1}{\Delta V} \sum_k \left\{ \sum_i \left[ e_{i\alpha} e_{i\beta} f_i^{eq} + \left(1 - \frac{1}{2\tau_f}\right) e_{i\alpha} e_{i\beta} f_i^{(1)} \right] \right\} \Delta S_{k\beta} = 0 \tag{27}$$

$$\partial_t (\rho c_p T) + \frac{1}{\Delta V} \sum_k \left\{ \sum_i \left[ e_{i\alpha} g_i^{eq} + \left(1 - \frac{1}{2\tau_g}\right) e_{i\alpha} g_i^{(1)} \right] \right\} \Delta S_{k\beta} = 0 \tag{28}$$

where $\Delta V$ denotes the control volume, subscript $k$ denotes the $k^{\text{th}}$ interface of the control volume, $\Delta S_k$ denotes the area of the $k^{\text{th}}$ interface.

To calculate the interface fluxes, a local thermal lattice Boltzmann model is reconstructed at the interface. As shown in Fig. 1, a unit lattice of the D2Q9 model is constructed at the interface, $\mathbf{x}_S$, $\mathbf{x}_L$, and $\mathbf{x}_R$ are the positions of the interface center, left cell center, and right cell center, respectively. Note that for the 3D situation, a unit lattice of the D3Q19 model is reconstructed at the interface center. The key point of calculating interface fluxes is to evaluate the non-equilibrium density distribution function $f_i^{(1)}$ and the non-equilibrium energy distribution function $g_i^{(1)}$. The corresponding processes are summarized as follows:

(a) Calculation of $\rho$, $u_\alpha$ and $T$ at $\mathbf{x}_S - \mathbf{e}_i \delta t$ through interpolation. The interpolation scheme is given as

$$\phi = \begin{cases} \phi_L + \nabla \phi_L \cdot (\mathbf{x}_S - \mathbf{e}_i \delta t - \mathbf{x}_L) & \mathbf{x}_S - \mathbf{e}_i \delta t \text{ at the left cell} \\ \phi_R + \nabla \phi_R \cdot (\mathbf{x}_S - \mathbf{e}_i \delta t - \mathbf{x}_R) & \mathbf{x}_S - \mathbf{e}_i \delta t \text{ at the right cell} \\ 0.5 \left[ \phi_L + \nabla \phi_L \cdot (\mathbf{x}_S - \mathbf{e}_i \delta t - \mathbf{x}_L) + \phi_R + \nabla \phi_R \cdot (\mathbf{x}_S - \mathbf{e}_i \delta t - \mathbf{x}_R) \right] & \mathbf{x}_S - \mathbf{e}_i \delta t \text{ at the interface} \end{cases} \tag{29}$$

where $\phi = \rho$, $u_\alpha$ or $T$, $\mathbf{x}_L$ and $\mathbf{x}_R$ are the centers of the left and right cells, respectively. Using the macroscopic variables at $\mathbf{x}_S - \mathbf{e}_i \delta t$, $f_i^{eq}(\mathbf{x}_S - \mathbf{e}_i \delta t, t)$ and $g_i^{eq}(\mathbf{x}_S - \mathbf{e}_i \delta t, t)$ can be obtained by Eqs. (3) and (4), respectively.

(b) Time evolution of the thermal lattice Boltzmann model. The predicted macroscopic variables at $\mathbf{x}_S$ and $t + \delta t$ are calculated by

$$\rho^*(\mathbf{x}_s, t + \delta t) = \sum_i f_i^{eq}(\mathbf{x}_s - \mathbf{e}_i \delta t, t), \tag{30}$$

$$\rho^* u_\alpha^*(\mathbf{x}_s, t + \delta t) = \sum_i e_{i\alpha} f_i^{eq}(\mathbf{x}_s - \mathbf{e}_i \delta t, t), \tag{31}$$

$$\rho^* c_p T^*(\mathbf{x}_s, t + \delta t) = \sum_i g_i^{eq}(\mathbf{x}_s - \mathbf{e}_i \delta t, t), \tag{32}$$

where the superscript * denotes the predicted variables. Then, $f_i^{eq}(\mathbf{x}_S, t + \delta t)$ and $g_i^{eq}(\mathbf{x}_S, t + \delta t)$ are respectively determined by Eqs. (3) and (4) with the predicted variables.

(c) Calculation of the interface fluxes. In Eqs. (23), (24), and (25), the equilibrium distribution functions $f_i^{eq}$ and $g_i^{eq}$ are replaced by $f_i^{eq}(\mathbf{x}_S, t + \delta t)$ and $g_i^{eq}(\mathbf{x}_S, t+\delta t)$, respectively. According to Eqs. (1) and (2), the non-equilibrium distribution functions $f_i^{(1)}$ and $g_i^{(1)}$ can be expressed as

$$f_i^{(1)} = -\tau_f \delta t (\partial_t + e_{i\alpha}\partial_{1\alpha}) f_i^{eq} + O(\delta t^2)$$
$$= -\tau_f \left[ f_i^{eq}(\mathbf{x}_S, t+\delta t) - f_i^{eq}(\mathbf{x}_S - \mathbf{e}_i\delta t, t) \right] + O(\delta t^2), \quad (33)$$

$$g_i^{(1)} = -\tau_g \delta t (\partial_t + e_{i\alpha}\partial_{1\alpha}) g_i^{eq} + O(\delta t^2)$$
$$= -\tau_g \left[ g_i^{eq}(\mathbf{x}_S, t+\delta t) - g_i^{eq}(\mathbf{x}_S - \mathbf{e}_i\delta t, t) \right] + O(\delta t^2). \quad (34)$$

Finally, the formulations of TLBFS can be given as

$$\partial_t \rho = -\frac{1}{\Delta V} \sum_k P_\alpha \Delta S_{k\alpha}, \quad (35)$$

$$\partial_t (\rho u_\alpha) = -\frac{1}{\Delta V} \sum_k \Pi_{\alpha\beta} \Delta S_{k\beta}, \quad (36)$$

$$\partial_t (\rho c_p T) = -\frac{1}{\Delta V} \sum_k Q_\alpha \Delta S_{k\alpha}, \quad (37)$$

$$P_\alpha = \sum_i e_{i\alpha} f_i^{eq}(\mathbf{x}_S, t+\delta t), \quad (38)$$

$$\Pi_{\alpha\beta} = \sum_i \left\{ e_{i\alpha}e_{i\beta} f_i^{eq}(\mathbf{x}_S, t+\delta t) + (0.5-\tau_f) e_{i\alpha}e_{i\beta} \left[ f_i^{eq}(\mathbf{x}_S, t+\delta t) - f_i^{eq}(\mathbf{x}_S - \mathbf{e}_i\delta t, t) \right] \right\}, \quad (39)$$

$$Q_\alpha = \sum_i \left\{ e_{i\alpha} g_i^{eq}(\mathbf{x}_S, t+\delta t) + (0.5-\tau_g) e_{i\alpha} \left[ g_i^{eq}(\mathbf{x}_S, t+\delta t) - g_i^{eq}(\mathbf{x}_S - \mathbf{e}_i\delta t, t) \right] \right\}. \quad (40)$$

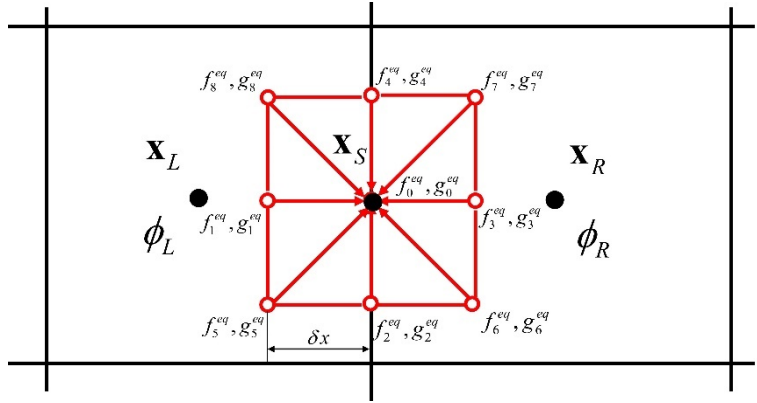

**Figure 1.** The reconstructed unit lattice of TLBFS.

In the present study, to illustrate the mechanism of good numerical stability of TLBFS, the first-order explicit scheme is adopted for the time discretization of TLBFS. The detailed formulations are

$$\rho^{n+1} = \rho^n - \frac{\Delta t}{\Delta V} \sum_k P_\alpha \Delta S_{k\alpha}, \quad (41)$$

$$(\rho u_\alpha)^{n+1} = (\rho u_\alpha)^n - \frac{\Delta t}{\Delta V} \sum_k \Pi_{\alpha\beta} \Delta S_{k\beta}, \quad (42)$$

$$(\rho c_p T)^{n+1} = (\rho c_p T)^n - \frac{\Delta t}{\Delta V} \sum_k Q_\alpha \Delta S_{k\alpha}. \quad (43)$$

where $\Delta t$ is the macroscopic time interval, the superscripts $n$ and $n+1$ denote the current and next time steps, respectively.

## 3. Derivation and analysis of MEs-TLBFS with actual numerical dissipation terms

In this section, MEs-TLBFS with actual numerical dissipation terms are derived by approximating the actual computational process. Through analyzing the discretized MEs-TLBFS, the mechanisms of the good performance of TLBFS can be revealed.

### 3.1 MEs-TLBFS with actual numerical dissipation terms

By using the second-order Taylor series expansion, $f_i^{eq}(\mathbf{x}_S - \mathbf{e}_i \delta t, t)$ and $g_i^{eq}(\mathbf{x}_S - \mathbf{e}_i \delta t, t)$ can be expanded as

$$f_i^{eq}(\mathbf{x}_S - \mathbf{e}_i \delta t, t) = f_i^{eq} - e_{i\alpha} \delta t \partial_\alpha f_i^{eq} + 0.5 \delta t^2 e_{i\alpha} e_{i\beta} \partial_\alpha \partial_\beta f_i^{eq} + O(\delta t^3), \quad (44)$$

$$g_i^{eq}(\mathbf{x}_S - \mathbf{e}_i \delta t, t) = g_i^{eq} - e_{i\alpha} \delta t \partial_\alpha g_i^{eq} + 0.5 \delta t^2 e_{i\alpha} e_{i\beta} \partial_\alpha \partial_\beta g_i^{eq} + O(\delta t^3). \quad (45)$$

Substituting Eqs. (7) and (44) into Eqs. (30) and (31), the macroscopic density and momentum equations recovered from the local thermal lattice Boltzmann model are

$$\rho^* = \rho^n - \partial_\alpha (\rho u_\alpha)^n \delta t + \frac{1}{2} \delta t^2 \partial_\alpha \partial_\beta (\rho u_\alpha u_\beta + \rho c_s^2 \delta_{\alpha\beta})^n + O(\delta t^3), \quad (46)$$

$$\begin{aligned}(\rho u_\alpha)^* &= (\rho u_\alpha)^n - \delta t \partial_\beta (\rho u_\alpha u_\beta + \rho c_s^2 \delta_{\alpha\beta})^n \\ &\quad + 0.5 c_s^2 \delta t^2 \partial_\beta \left[ \partial_\beta (\rho u_\alpha) + \partial_\alpha (\rho u_\beta) + \partial_\gamma (\rho u_\gamma) \delta_{\alpha\beta} \right]^n + O(\delta t^3) \end{aligned}. \quad (47)$$

Substituting Eqs. (8) and (45) into Eq. (32), the macroscopic energy equation recovered from the local thermal lattice Boltzmann model is

$$(\rho c_p T)^* = (\rho c_p T)^n - \delta t \partial_\alpha (\rho c_p u_\alpha T)^n + 0.5 c_s^2 \delta t^2 \partial_\alpha \partial_\beta (\rho c_p T \delta_{\alpha\beta})^n + O(\delta t^3). \quad (48)$$

Then, by combining with Eqs. (7) and (8), the interface fluxes of TLBFS, i.e., Eqs. (38), (39), and (40), can be simplified as

$$P_\alpha = (\rho u_\alpha)^*, \quad (49)$$

$$\begin{aligned}\Pi_{\alpha\beta} &= e_{i\alpha} e_{i\beta} f_i^{eq} + (0.5 - \tau_f) e_{i\alpha} e_{i\beta} \begin{bmatrix} f_i^{eq}(\mathbf{x}_S, t+\delta t) - f_i^{eq}(\mathbf{x}_S, t) \\ + e_{i\gamma} \delta t \partial_\gamma f_i^{eq}(\mathbf{x}_S, t) + O(\delta t^2) \end{bmatrix} \\ &= (\rho u_\alpha u_\beta + \rho c_s^2 \delta_{\alpha\beta})^* - \upsilon \left[ \partial_\beta (\rho u_\alpha) + \partial_\alpha (\rho u_\beta) + \partial_\gamma (\rho u_\gamma) \delta_{\alpha\beta} \right] \\ &\quad - (\tau_f - 0.5) \left[ (\rho u_\alpha u_\beta + \rho c_s^2 \delta_{\alpha\beta})^* - (\rho u_\alpha u_\beta + \rho c_s^2 \delta_{\alpha\beta}) \right] + O(\delta t^2) \end{aligned}, \quad (50)$$

$$\begin{aligned}Q_\alpha &= \sum_i e_{i\alpha} g_i^{eq}(\mathbf{x}_S, t+\delta t) + (0.5 - \tau_g) \sum_i e_{i\alpha} \begin{bmatrix} g_i^{eq}(\mathbf{x}_S, t+\delta t) - g_i^{eq}(\mathbf{x}_S, t) \\ + e_{i\beta} \delta t \partial_\beta g_i^{eq}(\mathbf{x}_S, t) + O(\delta t^2) \end{bmatrix} \\ &= (\rho c_p u_\alpha T)^* - \chi \partial_\alpha (\rho c_p T) - (\tau_g - 0.5) \left[ (\rho c_p u_\alpha T)^* - (\rho c_p u_\alpha T) \right] + O(\delta t^2) \end{aligned}, \quad (51)$$

where $\upsilon=(\tau_f-0.5)c_s^2\delta t$ and $\chi=(\tau_g-0.5)c_s^2\delta t$.

Substituting Eqs. (39) and (50) into Eqs. (35) and (36), we can obtain the macroscopic density and momentum equations of TLBFS as follows:

$$\partial_t\rho = -\partial_\alpha(\rho u_\alpha) - \delta t \partial_\alpha \partial_\beta(\rho u_\alpha u_\beta + \rho c_s^2 \delta_{\alpha\beta}) \\ + 0.5 c_s^2 \delta t^2 \partial_\alpha \partial_\beta [\partial_\beta(\rho u_\alpha) + \partial_\alpha(\rho u_\beta) + \partial_\gamma(\rho u_\gamma)\delta_{\alpha\beta}],\quad (52)$$

$$\partial_t(\rho u_\alpha) = -\partial_\beta(\rho u_\alpha u_\beta + \rho c_s^2 \delta_{\alpha\beta}) + \upsilon \partial_\beta[\partial_\beta(\rho u_\alpha) + \partial_\beta(\rho u_\alpha) + \partial_\gamma(\rho u_\gamma)\delta_{\alpha\beta}] \\ + (\tau_f - 1.5)[\partial_\beta(\rho u_\alpha u_\beta + \rho c_s^2 \delta_{\alpha\beta})^* - \partial_\beta(\rho u_\alpha u_\beta + \rho c_s^2 \delta_{\alpha\beta})].\quad (53)$$

Similarly, substituting Eq. (51) into Eq. (37), the macroscopic energy equation of TLBFS can be given as follows:

$$\partial_t(\rho c_p T) = -\partial_\alpha(\rho c_p u_\alpha T) + \chi \partial_\alpha \partial_\beta(\rho c_p T \delta_{\alpha\beta}) \\ + (\tau_g - 1.5)[\partial_\alpha(\rho c_p u_\alpha T)^* - \partial_\alpha(\rho c_p u_\alpha T)].\quad (54)$$

In the present study, Eqs. (46), (47), (48), (52), (53) and (54) are labeled by MEs-TLBFS.

### 3.2 Discretization of MEs-TLBFS with FVM

By directly solving MEs-TLBFS with FVM, TLBFS can be reconstructed with macroscopic variables. The reconstructed TLBFS is labeled by RTLBFS. It should be noted that TLBFS uses the staggered mesh where the predicted variables and the updated variables locate at the interface centers and cell centers, respectively. It has been proven that the staggered mesh is essential to restrain pressure oscillation. Therefore, in the discretized MEs-TLBFS, the staggered mesh is adopted as well. Referring to the discretization scheme of TLBFS, MEs-TLBFS are solved through two steps as well.

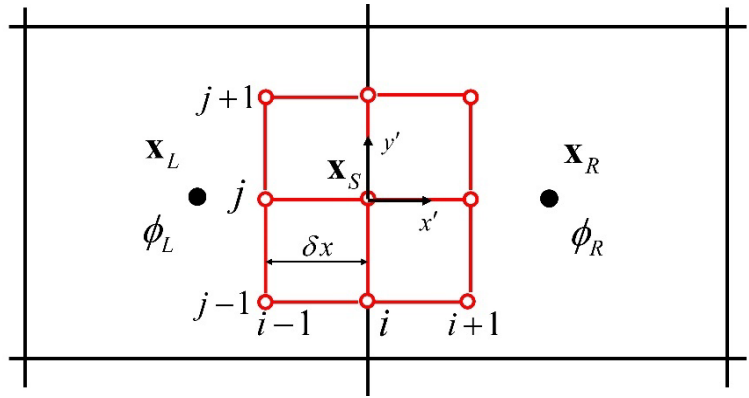

**Figure 2.** The reconstructed unit lattice at the interface.

(a) Predictor step. Firstly, similar to TLBFS, a unit lattice illustrated in Fig. 2 is reconstructed as well. Note that the local coordinate is adopted for the reconstructed unit lattice to maintain the lattice symmetric about the interface. Secondly, the variables at the 9 lattice nodes are interpolated by Eq. (29). The partial derivatives at cell centers are calculated by the discretized Gauss theorem as follows:

$$\partial_\alpha \phi = \sum_k \phi_s \Delta S_{k\alpha} . \qquad (55)$$

The macroscopic variables at the interface in Eq. (55) are interpolated by

$$\phi_S = \frac{\phi_L |\mathbf{x}_R - \mathbf{x}_s| + \phi_R |\mathbf{x}_L - \mathbf{x}_s|}{|\mathbf{x}_R - \mathbf{x}_s| + |\mathbf{x}_L - \mathbf{x}_s|}, \qquad (56)$$

where $\phi$ denotes an arbitrary variable, the subscripts $S$, $L$, and $R$ denote the variables at the interface, the left cell, and the right cell, respectively. Thirdly, Eqs. (46), (47), and (48) are discretized by the finite difference method to obtain the predicted variables at the interface. The general 3D discretization scheme of these equations can be found in Ref. [27].

(b) Corrector step. The updated variables $\rho^{n+1}$, $(\rho u_\alpha)^{n+1}$ and $(\rho c_p T)^{n+1}$ at cell centers can be obtained by

$$\rho^{n+1} = \rho^n - \frac{\Delta t}{\Delta V} \sum_k (\rho u_\alpha)^* \Delta S_{k\alpha}, \qquad (57)$$

$$(\rho u_\alpha)^{n+1} = (\rho u_\alpha)^n + \frac{\Delta t}{\Delta V} \sum_k \left\{ \begin{array}{l} -(\rho u_\alpha u_\beta + \rho c_s^2 \delta_{\alpha\beta})^n \\ +\upsilon \left[ \partial_\beta (\rho u_\alpha) + \partial_\alpha (\rho u_\beta) + \partial_\gamma (\rho u_\gamma) \delta_{\alpha\beta} \right]^n \end{array} \right\} \Delta S_{k\beta}$$
$$+ \frac{(\tau_f - 1.5) \Delta t}{\Delta V} \sum_k \left[ (\rho u_\alpha u_\beta + \rho c_s^2 \delta_{\alpha\beta})^* - (\rho u_\alpha u_\beta + \rho c_s^2 \delta_{\alpha\beta})^n \right] \Delta S_{k\beta} \qquad (58)$$

$$(\rho c_p T)^{n+1} = (\rho c_p T)^n + \frac{\Delta t}{\Delta V} \sum_k \left\{ -(\rho c_p u_\alpha T)^n + \chi \partial_\alpha (\rho c_p T)^n \right\} \Delta S_{k\alpha}$$
$$+ \frac{(\tau_g - 1.5) \Delta t}{\Delta V} \sum_k \left[ (\rho c_p u_\alpha T)^* - (\rho c_p u_\alpha T)^n \right] \Delta S_{k\alpha} \qquad (59)$$

**3.3 The mechanism of the good numerical stability of TLBFS for high Rayleigh number flows**

The numerical dissipation terms are essential to stabilize the computation of weakly compressible models. Since TLBFS belongs to the weakly compressible model, TLBFS needs to introduce some numerical dissipation terms to stabilize computation as well. Consequently, this section analyzes the effect of numerical dissipation terms on the numerical stability of TLBFS.

It can be seen from MEs-TLBFS that for $\delta t = 0$, MEs-TLBFS degenerate to

$$\partial_t \rho = -\partial_\alpha (\rho u_\alpha), \qquad (60)$$

$$\partial_t (\rho u_\alpha) = -\partial_\beta (\rho u_\alpha u_\beta + \rho c_s^2 \delta_{\alpha\beta}) + \mu \partial_\beta \left[ \partial_\alpha u_\beta + \partial_\beta u_\alpha + u_\gamma \delta_{\alpha\beta} \right]$$
$$+ \underbrace{\upsilon \partial_\beta \left[ u_\alpha \partial_\beta \rho + u_\beta \partial_\alpha \rho + u_\gamma \partial_\gamma \rho \delta_{\alpha\beta} \right]}_{o(Ma^3)}, \qquad (61)$$

$$\partial_t (\rho c_p T) = -\partial_\alpha (\rho c_p u_\alpha T) + \lambda \partial_\alpha \partial_\beta (T \delta_{\alpha\beta}) + \underbrace{\chi c_p T \partial_\alpha \partial_\beta (\rho \delta_{\alpha\beta})}_{o(Ma^2)}, \qquad (62)$$

which are consistent with the standard macroscopic governing equations with some extra higher-order terms. When $\delta t \neq 0$, some additional small terms exist in MEs-TLBFS, i.e., $\Delta t \delta t \partial_\alpha \partial_\beta \left( \rho u_\alpha u_\beta + \rho c_s^2 \delta_{\alpha\beta} \right)^n + 1.5 c_s^2 \delta t^2 \Delta t \partial_\beta \partial_\beta \partial_\alpha \left( \rho u_\alpha \right)^n$ in the density equation, $(\tau_f - 1.5)\Delta t \left[ \partial_\beta \left( \rho u_\alpha u_\beta + \rho c_s^2 \delta_{\alpha\beta} \right)^* - \partial_\beta \left( \rho u_\alpha u_\beta + \rho c_s^2 \delta_{\alpha\beta} \right)^n \right]$ in the momentum equation, $(\tau_g - 1.5)\Delta t \left[ \partial_\alpha \left( \rho c_p u_\alpha T \right)^* - \partial_\alpha \left( \rho c_p u_\alpha T \right)^n \right]$ in the energy equation.

To investigate the effect of these additional terms on the numerical stability of RTLBFS and TLBFS, 2D natural convection in a square cavity is simulated. As shown in Fig. 3, the cavity size is $L \times L$, the four walls of the square cavity are stationary. The temperatures of the left and right walls are fixed at $T_h$ and $T_l$ ($T_h > T_l$), respectively, while the upper and lower walls are adiabatic. The problem is characterized by the Rayleigh number Ra and the Prandtl number Pr, which are defined as

$$\text{Ra} = \frac{\beta g (T_h - T_l) L^3}{\upsilon \chi} = \frac{V_c^2 L^2}{\chi \upsilon}, \quad \text{Pr} = \frac{\upsilon}{\chi}, \tag{63}$$

where $V_c = \sqrt{\beta g (T_h - T_l) L}$ is the characteristic velocity constrained by the low Mach number limit. In the present test, uniform meshes are adopted, $\text{Pr} = 0.71$ and $V_c = 0.1$ are chosen. The Boussinesq approximation is adopted and the buoyancy can be given as [23]

$$\mathbf{F} = \beta \mathbf{g} \rho (T - T_{ref}), \quad \mathbf{g} = (0, -g) \tag{64}$$

where $T_{ref}$ is the reference temperature set as $(T_h + T_l)/2$.

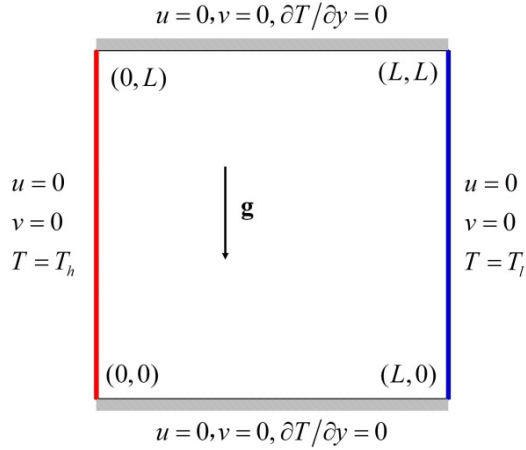

**Figure 3.** Schematic of 2D lid-driven cavity flow.

Firstly, the convergence situations of RTLBFS and TLBFS at different Ra and $\delta t$ are investigated. The mesh used for the test is uniform and its size is fixed at 50×50. Other parameters are set as $\Delta t = 0.5$, $\Delta x = 1$, $c_s^2 = 1/3$. The convergence criterion is

$$\frac{\sum \left| |\mathbf{u}|^{n+1} - |\mathbf{u}|^n \right|}{\sum |\mathbf{u}|^{n+1}} < 10^{-9}, \quad \frac{\sum \left| T^{n+1} - T^n \right|}{\sum T^{n+1}} < 10^{-9}. \tag{65}$$

As shown in Fig. 4, when $\delta t$ is small, i.e., the additional terms are relatively small, both RTLBFS and TLBFS are easy to diverge. With $\delta t$ increasing, i.e., the additional terms become relatively larger, the numerical stability of RTLBFS and TLBFS shows obvious improvement. The result proves that those additional terms include some numerical dissipation terms which are essential to stabilize computation. Besides, the stable parameter ranges of RTLBFS and TLBFS show good agreement, which indicates that RTLBFS can recover the stability of TLBFS well.

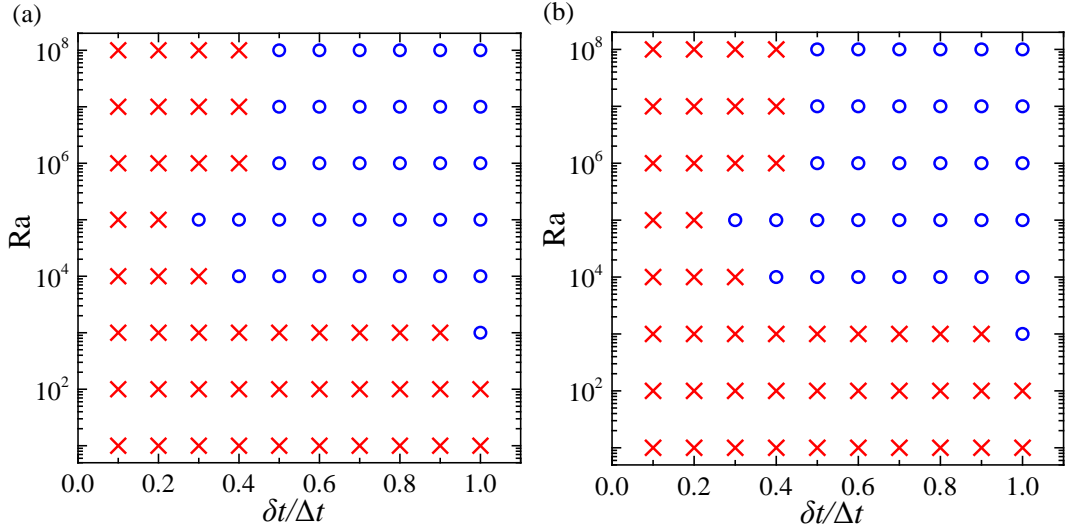

**Figure 4.** Convergence situations of RTLBFS (a) and TLBFS (b). The blue circle indicates convergence while the red cross indicates divergence.

Secondly, the convergence situations of RTLBFS and TLBFS at different Ra and $N$ are investigated, where $N$ is the mesh number in length $L$. The fixed parameters are $\delta t = 1$, $V_c = 0.1$, $c_s^2 = 1/3$. As shown in Fig. 5, both of them have good numerical stability at high Ra. For $\mathrm{Ra} = 10^8$, their results are still stable even for a coarse mesh size 10×10. Besides, the stable parameter ranges of RTLBFS and TLBFS show good agreement again. It indicates that RTLBFS preserves the mechanism of the good numerical stability of TLBFS for high Rayleigh number thermal flows very well.

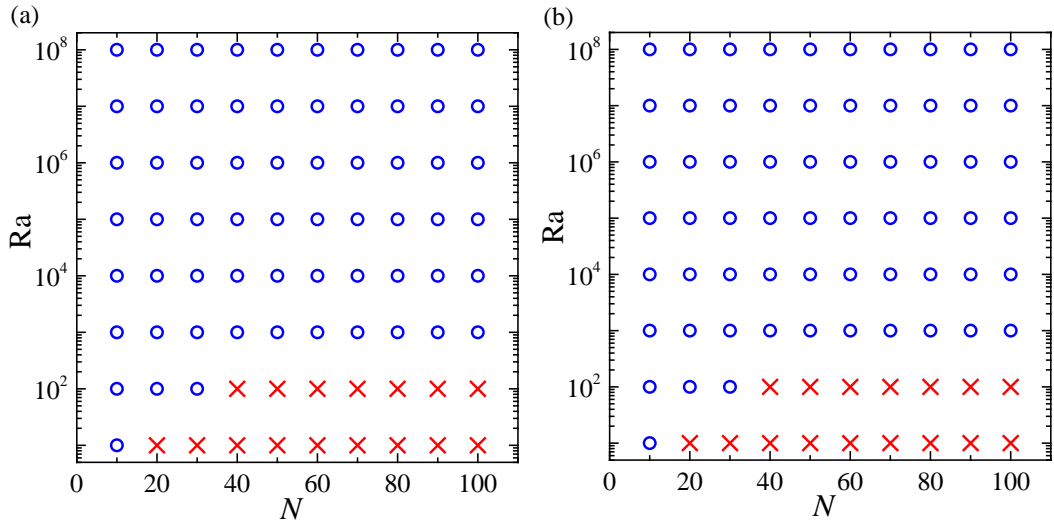

**Figure 5.** Convergence situations of the RTLBFS (a) and TLBFS (b) at different Ra and $N$. The blue circle indicates convergence while the red cross indicates divergence.

## 4. Numerical tests

To validate RTLBFS, several benchmarks are simulated in this section.

### 4.1. Transient heat diffusion of a Gaussian hill

Transient heat diffusion of a Gaussian hill is simulated in the present subsection to compare the accuracy of RTLBFS and TLBFS. The computational domain is set as $[0,2L] \times [0,2L]$ and has a constant uniform velocity $\mathbf{u} = (u_0, 0)$. The initial temperature distribution is a Gaussian distribution

$$T(\mathbf{x},0) = \frac{T_0}{2\pi\sigma^2} \exp\left[\frac{-(\mathbf{x}-\mathbf{x}_0)^2}{2\sigma^2}\right], \tag{66}$$

where $\sigma$ is a constant set as $\sigma = 0.05L$, $T_0$ is the reference temperature set as $T_0 = 2\pi\sigma^2$, $\mathbf{x}_0$ is the position of the center of the Gaussian hill set as $\mathbf{x}_0 = (0.5L, 0)$. The analytical solution is

$$T(\mathbf{x},t) = \frac{T_0}{2\pi(2\chi t + \sigma^2)} \exp\left[\frac{-(\mathbf{x}-\mathbf{x}_0-\mathbf{u}t)^2}{2(2\chi t + \sigma^2)}\right]. \tag{67}$$

The periodic boundary is imposed at the four boundaries and the simulation time is set to be short enough to avoid obvious fluctuation at the boundaries.

Firstly, to validate RTLBFS, the temperature contour at Fourier number $\mathrm{Fo} = t\chi/L^2 = 0.005$ is compared with the corresponding analytical solutions. Uniform mesh is adopted and the parameters are set as follows: $\delta t = 0.5$, $\Delta t = 1$, the mesh spacing $\Delta x = L/50$, the Peclect number $\mathrm{Pe} = u_0 L/\chi = 100$, the mesh Fournier number $\mathrm{Fo}_{\Delta x} = \Delta t \chi/\Delta x^2 = 0.1$. As shown in Fig. 6, the results obtained by RTLBFS show good agreement with the analytical solution.

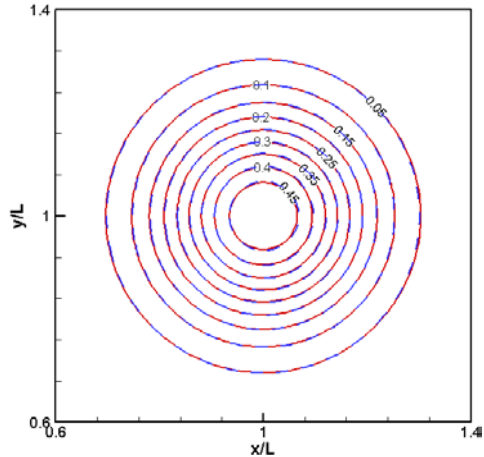

**Figure 6.** Comparison of temperature contours at Fo=0.005 between the calculated results (blue dash line) and analytical solution (red solid line).

To compare the accuracy of TLBFS and RTLBFS, the convergence orders of the two methods are compared. The fixed parameters include $\delta t = 0.5$, $\Delta t = 1$, $\text{Pe} = u_0 L/\chi = 10$, $\text{Fo}_{\Delta x} = \Delta t \chi / \Delta x^2 = 0.1$, while the mesh has 5 different sizes (80×80, 120×120, 160×160, 200×200, 240×240). The numerical error quantified by $L_2$ norm is defined as

$$E_2 = \sqrt{\frac{\sum_{i=1}^{N_0}(T-T_a)^2}{N_0}}, \qquad (68)$$

where $T_a$ is the analytical solution, $N_0$ is the grid amount. The numerical error is calculated at $\text{Fo} = t\chi/L^2 = 0.005$.

Fig. 7 shows that both TLBFS and RTLBFS have a convergence order near 2, which is consistent with the accuracy of the discretization scheme. Besides, the numerical error of RTLBFS is only slightly higher than that of TLBFS. Since RTLBFS is the reconstruction of TLBFS with second-order accuracy, the slight deviation of numerical error results from the different higher-order terms. The result indicates that RTLBFS can recover the accuracy of TLBFS well.

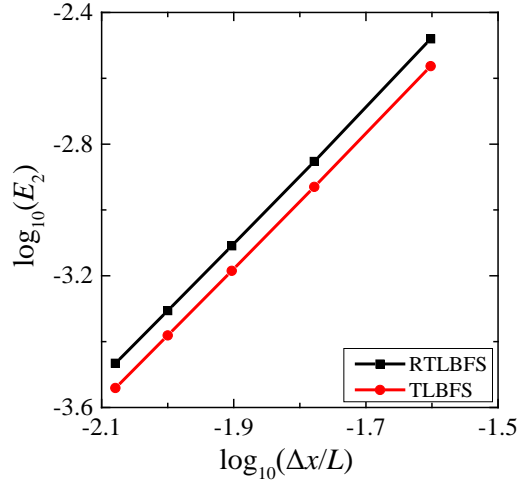

**Figure 7.** Convergence order analysis of TLBFS and RTLBFS for unsteady heat diffusion of a Gaussian hill.

### 4.2. 2D natural convection in a square cavity

To validate RTLBFS for simulating natural convection, 2D natural convection in a square cavity is simulated in the present subsection. The physical model has been described in Section 3.3. In the present subsection, the Prandtl number and characteristic velocity are set as Pr=0.71 and $V_c$=0.1, respectively, and six Rayleigh numbers ($10^3$, $10^4$, $10^5$, $10^6$, $10^7$, and $10^8$) are chosen. According to Yang et al. [32] and Kang and Hassan [33], uniform meshes of sizes 100×100, 150×150, 200×200, and 250×250 are adopted for Ra=$10^3$, $10^4$, $10^5$, and $10^6$, respectively. For Ra=$10^7$ and $10^8$, non-uniform meshes are adopted. The mesh sizes are set as 300×300 and 400×400, respectively, and the distances from the wall to the nearest computational grid are 0.0004$L$ and 0.0002$L$, respectively.

Firstly, the natural convection at Ra=$10^3$, $10^4$, $10^5$, and $10^6$ are investigated. The streamlines and isotherms computed by the RTLBFS are depicted in Figs. 8 and 9, respectively. These plots obtained by RTLBFS are in good accordance with those obtained by the thermal LBM (TLBM) [9], TLBFS [24], and the high order differential quadrature (DQ) method [34].

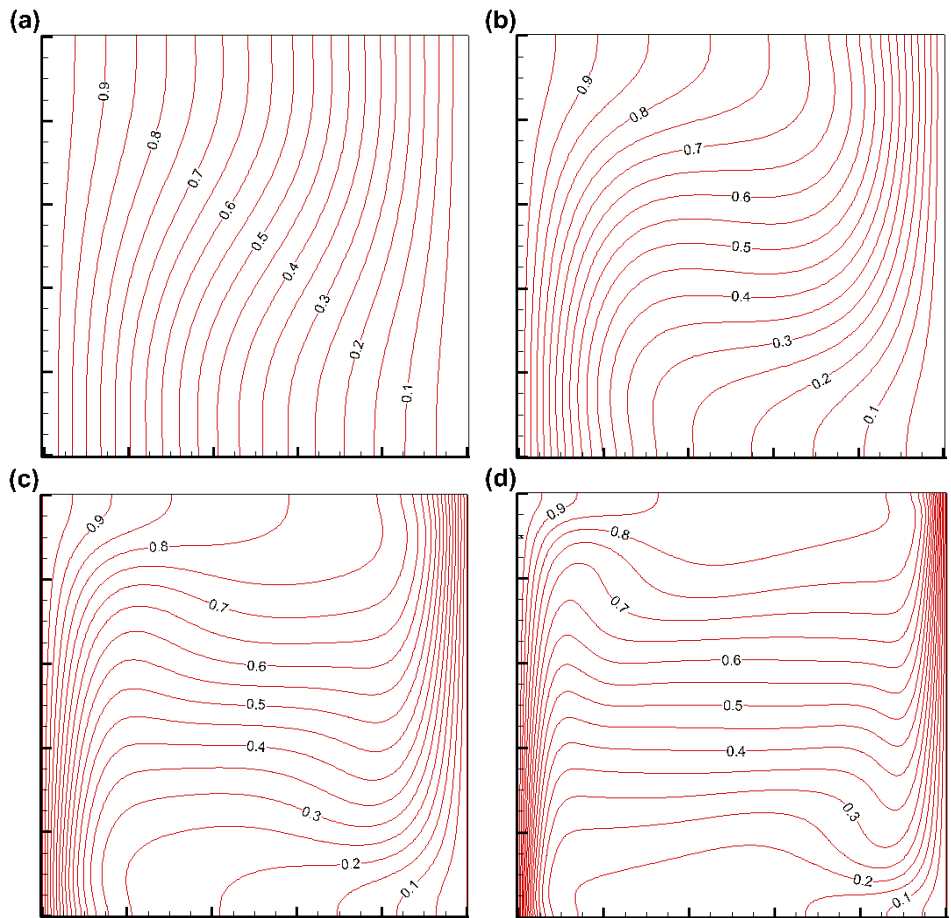

**Figure 8.** Isotherms for natural convection in a square cavity at four different Rayleigh numbers. (a) Ra=$10^3$, (b) Ra=$10^4$, (c) Ra=$10^5$, (d) Ra=$10^6$.

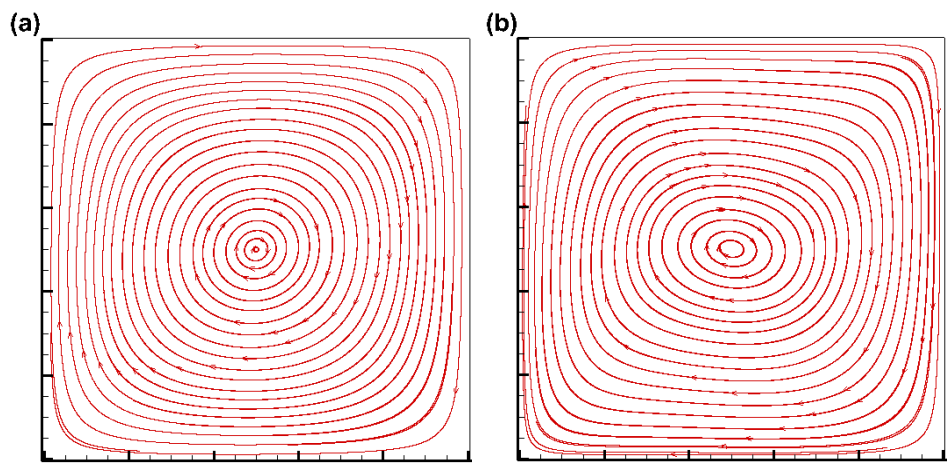

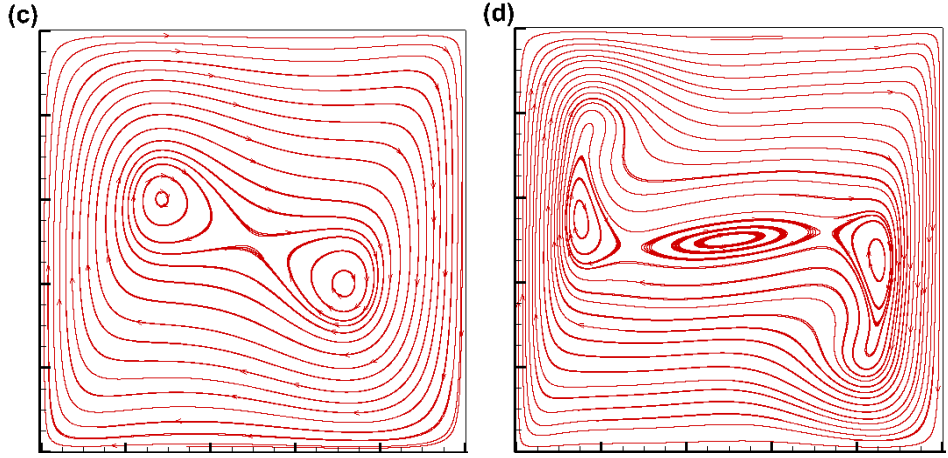

**Figure 9.** Streamlines for natural convection in a square cavity at four different Rayleigh numbers.
(a) Ra=$10^3$, (b) Ra=$10^4$, (c) Ra=$10^5$, (d) Ra=$10^6$.

To validate RTLBFS in quantity, Table 1 shows the comparison of several characteristic parameters, i.e., the maximum $u$ along $x=0.5L$ and the corresponding $y$ coordinate, the maximum $v$ along $y=0.5L$ and the corresponding $x$ coordinate, the average Nusselt number $\mathrm{Nu}_a$ of the computational domain. The average Nusselt number is defined as

$$\mathrm{Nu}_a = \frac{1}{\chi \Delta T L} \iint (uT - \chi \partial_x T) dxdy . \qquad (69)$$

**Table 1.** shows that these characteristic parameters at different Ra obtained by RTLBFS agree well with the reference results [9, 24, 34].

Table 1 Comparison of characteristic parameters for natural convection at different *Ra*.

| Ra | References | $10^3$ | $10^4$ | $10^5$ | $10^6$ |
|---|---|---|---|---|---|
| $(u_{x=0.5L}L/\chi)_{max}$ | TLBM [9] | 3.644 | 16.134 | 34.261 | 63.024 |
| | TLBFS [24] | 3.640 | 16.14 | 34.87 | 64.838 |
| | DQ [34] | 3.649 | 16.190 | 34.736 | 64.775 |
| | Present | 3.644 | 16.166 | 34.729 | 64.879 |
| $y/L$ | TLBM [9] | 0.810 | 0.820 | 0.855 | 0.848 |
| | TLBFS [24] | 0.815 | 0.825 | 0.855 | 0.850 |
| | DQ [34] | 0.815 | 0.825 | 0.855 | 0.850 |
| | Present | 0.814 | 0.824 | 0.855 | 0.850 |
| $(v_{y=0.5L}L/\chi)_{max}$ | TLBM [9] | 3.691 | 19.552 | 67.799 | 215.26 |
| | TLBFS [24] | 3.708 | 19.67 | 68.85 | 220.92 |
| | DQ [34] | 3.698 | 19.638 | 68.640 | 220.64 |
| | Present | 3.698 | 19.628 | 68.643 | 220.74 |
| $x/L$ | TLBM [9] | 0.180 | 0.120 | 0.065 | 0.040 |
| | TLBFS [24] | 0.180 | 0.118 | 0.065 | 0.038 |
| | DQ [34] | 0.180 | 0.120 | 0.065 | 0.035 |
| | Present | 0.180 | 0.118 | 0.066 | 0.038 |
| $\mathrm{Nu}_a$ | TLBM [9] | 1.117 | 2.241 | 4.511 | 8.731 |
| | TLBFS [24] | 1.115 | 2.232 | 4.491 | 8.711 |
| | DQ [34] | 1.118 | 2.245 | 4.523 | 8.762 |
| | Present | 1.118 | 2.245 | 4.522 | 8.832 |

Secondly, the test cases of high Ra ($10^7$ and $10^8$) are investigated. Figs. 10 and 11 show the corresponding isothermals and streamlines, respectively. Compared with the test cases at Ra=$10^3$ to $10^6$, the velocity and temperature boundary layers near the hot and cold walls are much thinner, which indicates strong natural convection with high heat transfer rates. Besides, the isothermals and streamlines given by RTLBFS match well with those given by Wang et al. [26] and Yang et al. [32].

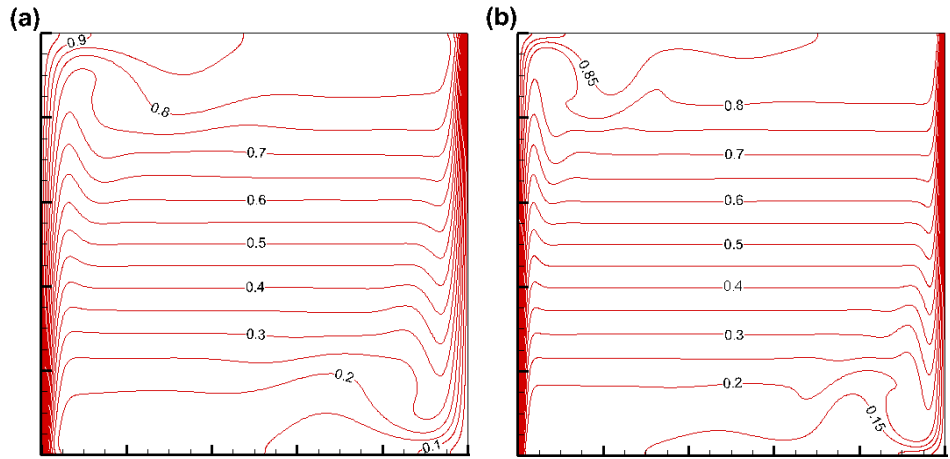

**Figure 10.** Isotherms for natural convection in a square cavity at (a) Ra=$10^7$, (b) Ra=$10^8$.

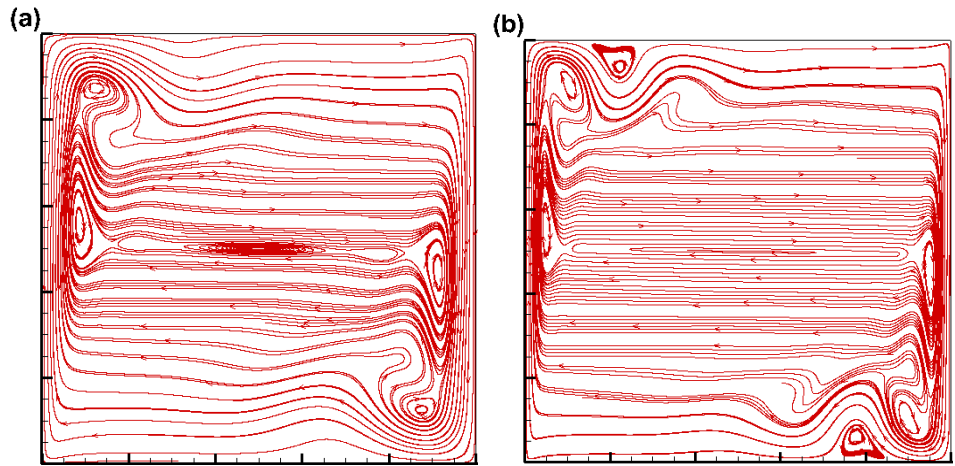

**Figure 11.** Streamlines for natural convection in a square cavity at (a) Ra=$10^7$, (b) Ra=$10^8$.

To validate the results in quantity, the *u* profiles along *y*=0.5*L* and *v* profiles along *x*=0.5*L* at Ra=$10^7$ and $10^8$ obtained by RTLBFS are compared with the reference results [26, 32]. As shown in Fig. 12, good agreement can be observed between the results given by RTLBFS and the reference results [26, 32].

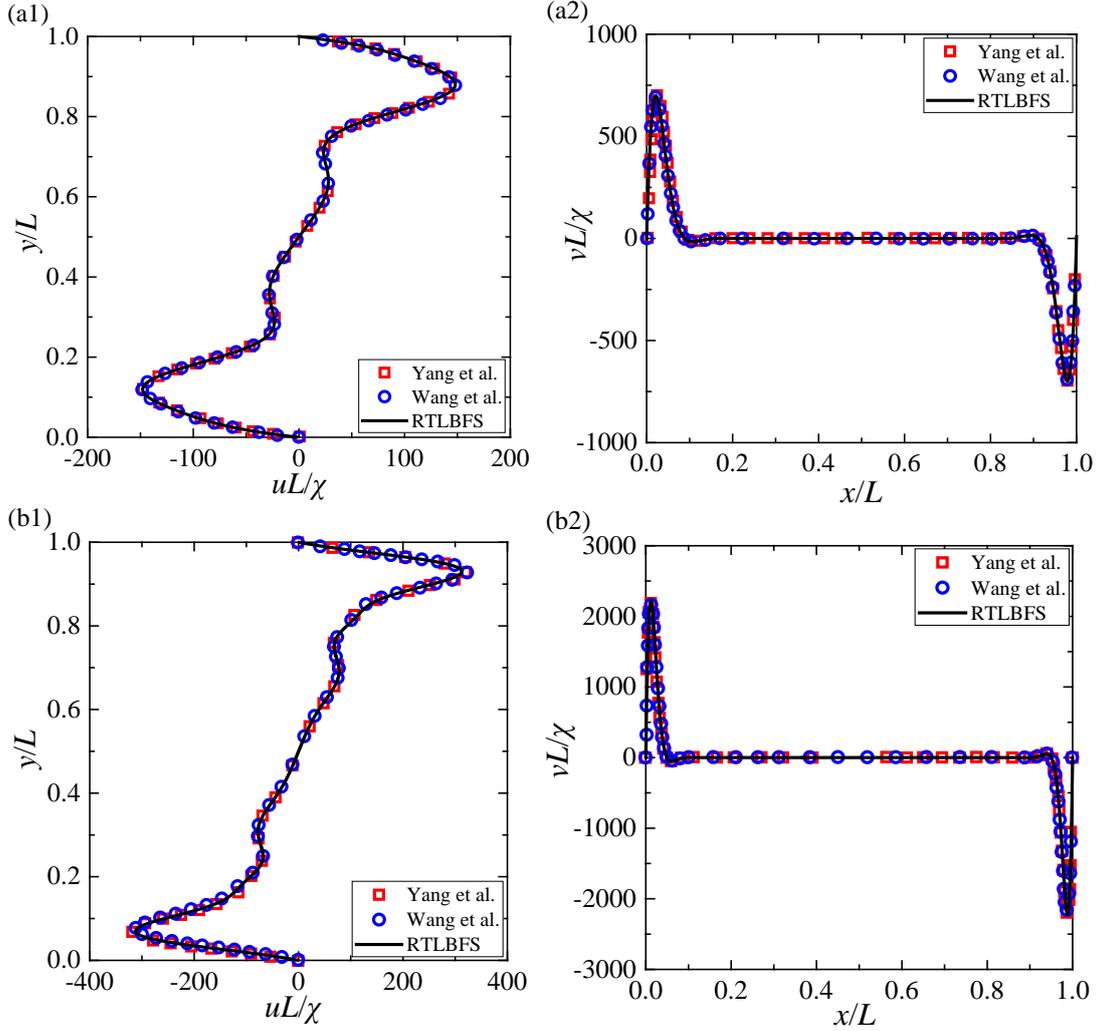

**Figure 12.** The *u* profiles along *x*=0.5*L* and *v* profiles along *y*=0.5*L* at Ra=$10^7$ (a1 and a2), Ra=$10^8$ (b1 and b2), together with those given by Yang et al. [32] and Wang et al. [24].

Besides, Tables 2 and 3 provide detailed comparisons of the maximum dimensionless streamfunction $|\psi/\chi|_{max}$ and the corresponding *x* and *y* coordinates, the average Nusselt number along *x*=0.5*L*, the maximum dimensionless horizontal velocity $(u_{x=0.5L}L/\chi)_{max}$ along *x*=0.5*L* and the corresponding *y* coordinate, the maximum dimensionless vertical velocity $(v_{y=0.5L}L/\chi)_{max}$ along *y*=0.5*L* and the corresponding *x* coordinates at Ra=$10^7$ and $10^8$, respectively. The reference results include simulation results of Contrino et al. [35] obtained by the multiple-relaxation-times thermal lattice Boltzmann equation, Quéré [36] calculated by the high order pseudospectral method, Yang et al. [32] calculated by a developed gas kinetic scheme, and Wang et al. [24] by TLBFS. Tables 2 and 3 demonstrate that the present RTLBFS results compare well with the reference data [24, 32, 35, 36].

**Table 2.** Results of 2D natural convection in a square cavity at Ra=$10^7$.

| Parameters | Contrino et al. [35] | Quéré [36] | Wang et al. [24] | Yang et al. [32] | Present |
|---|---|---|---|---|---|
| $|\psi/\chi|_{max}$ | 30.1760 | 30.165 | 30.164 | 30.192 | 30.121 |
| $x/L$ | 0.0858 | 0.086 | 0.0857 | 0.0848 | 0.0862 |
| $y/L$ | 0.5558 | 0.556 | 0.5559 | 0.5548 | 0.5584 |
| $\text{Nu}_{1/2}$ | 16.5231 | 16.52 | 16.543 | 16.518 | 16.510 |
| $(u_{x=0.5L}L/\chi)_{\max}$ | 148.5852 | 148.59 | 148.84 | 148.86 | 148.08 |
| $y/L$ | 0.8793 | 0.879 | 0.8789 | 0.8800 | 0.8802 |
| $(v_{y=0.5L}L/\chi)_{\max}$ | 699.3166 | 699.18 | 699.91 | 699.20 | 699.21 |
| $x/L$ | 0.0213 | 0.021 | 0.0216 | 0.0217 | 0.0213 |

**Table 3.** Results of 2D natural convection in a square cavity at Ra=$10^8$.

| Parameters | Contrino et al. [35] | Quéré [36] | Wang et al. [24] | Yang et al. [32] | Present |
|---|---|---|---|---|---|
| $\|\psi/\chi\|_{\max}$ | 53.953 | 53.85 | 53.893 | 53.885 | 53.823 |
| $x/L$ | 0.0480 | 0.048 | 0.0476 | 0.0476 | 0.0488 |
| $y/L$ | 0.5533 | 0.553 | 0.5528 | 0.5532 | 0.5516 |
| $\text{Nu}_{1/2}$ | 30.227 | 30.225 | 30.301 | 30.227 | 30.220 |
| $(u_{x=0.5L}L/\chi)_{\max}$ | 321.37 | 321.9 | 323.65 | 321.49 | 312.734 |
| $y/L$ | 0.9276 | 0.928 | 0.9288 | 0.9284 | 0.9303 |
| $(v_{y=0.5L}L/\chi)_{\max}$ | 2222.3 | 2222 | 2222.9 | 2221.7 | 2222.8 |
| $x/L$ | 0.0120 | 0.012 | 0.0119 | 0.0122 | 0.0118 |

### 4.3. Mixed convection from a heated circular cylinder

To validate RTLBFS for simulating mixed convection, the mixed heat transfer of a heated circular cylinder is simulated in this subsection. As shown in Fig. 13, the stationary circular cylinder of diameter $D$ has a uniform surface temperature $T_h = 1$. The free stream density, velocity, and temperature are $\rho_\infty = 1$, $u_\infty = 0.1c_s$ and $T_l = 0$, respectively. This mixed convection problem can be characterized by the Reynolds number Re, the Grashof number Gr, and the Prandtl number Pr. Pr is defined as Eq. (63), and Re, Gr, Ra are defined as

$$\text{Re} = u_\infty D/\upsilon, \tag{70}$$

$$\text{Gr} = \frac{\beta g (T_h - T_l) D^3}{\upsilon^2}, \tag{71}$$

$$\text{Ra} = \frac{\beta g (T_h - T_l) D^3}{\upsilon \chi} = \frac{V_c^2 L^2}{\chi \upsilon}. \tag{72}$$

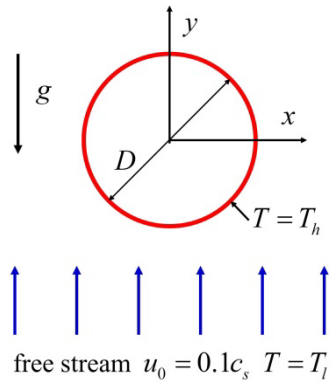

**Figure 13.** Physical model of mixed heat transfer from a heated circular cylinder.

The same as the simulations in Refs. [24] and [32], Re and Pr are fixed at Re=20 and Pr=0.7, while Gr has four values with Gr=0, 100, 800, and 1600. A body-fitted O-type mesh of size 300×450 is adopted for the four cases, and the far-field boundary is set at 52.5$D$ away from the center of the cylinder. The distance from the wall to the nearest computational grid is taken as 0.003$D$.

The computed streamlines at four different Gr are plotted in Fig. 14. For forced convection, i.e., Gr=0, there are two symmetric vortices behind the cylinder. With increasing Gr, the vortex length behind the cylinder decreases gradually until the vortex disappears for Gr≥800. The regularity is consistent with those observed by Wang et al. [24] and Yang et al. [32].

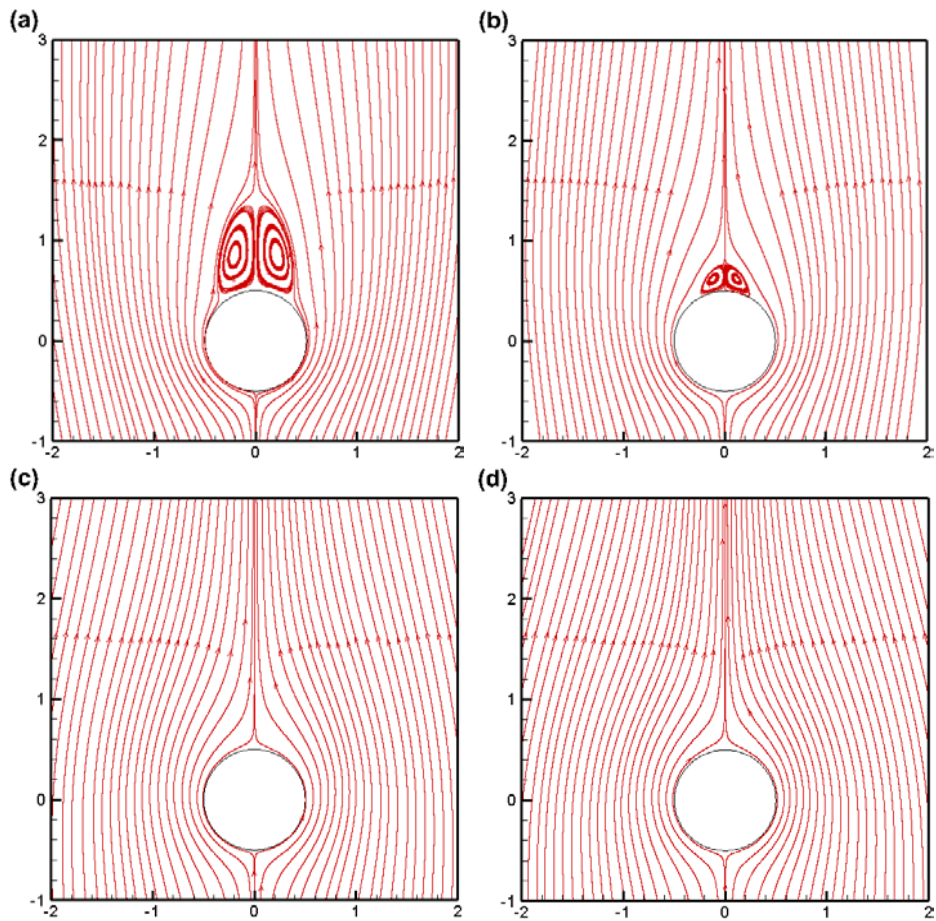

**Figure 14.** Streamlines for mixed convection at Re=20 and various Gr. (a) Gr=0, (b) Gr=100, (c) Gr=800, (d) Gr=1600.

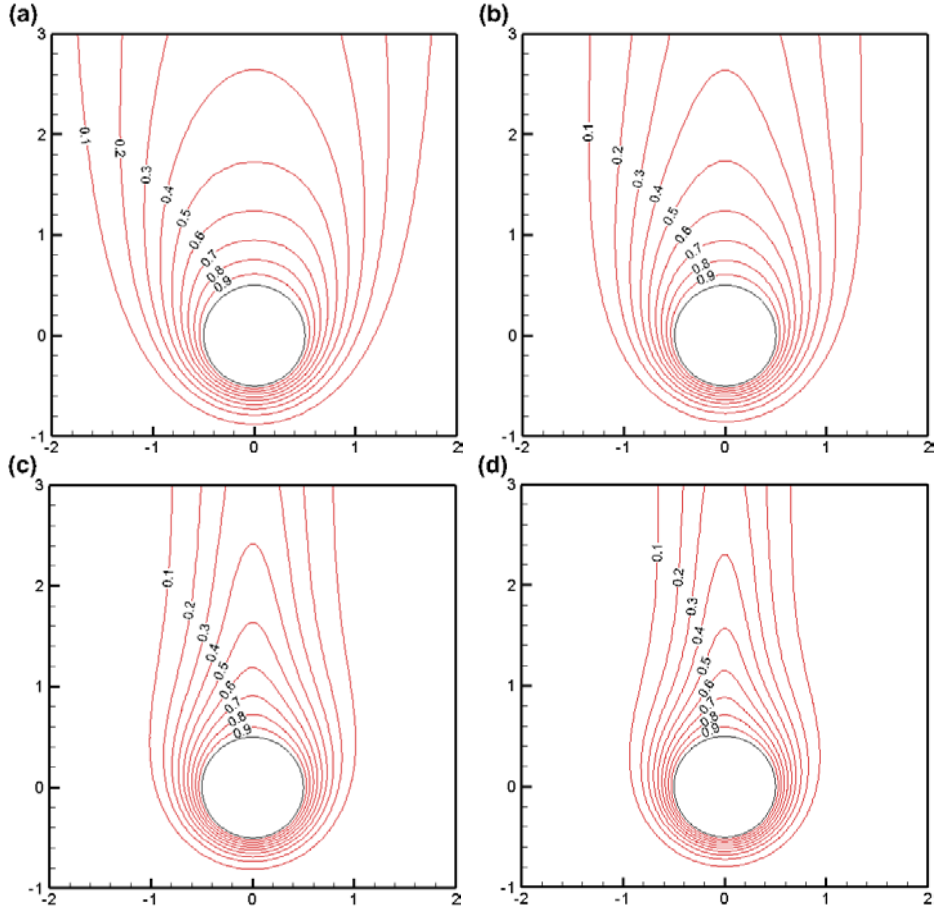

**Figure 15.** Isotherms for mixed convection at Re=20 and various Gr. (a) Gr=0, (b) Gr=100, (c) Gr=800, (d) Gr=1600.

Fig. 15 shows the computed isotherms at four different Gr. It can be seen that with increasing Gr, the isotherms near the cylinder surface become denser, which indicates that heat transfer is enhanced. Besides, for both streamlines and isotherms, the present RTLBFS results match well with those given by Wang et al. [24] and Yang et al. [32].

Furthermore, to make a quantitative comparison, the average Nusselt numbers $\overline{\mathrm{Nu}}$ and separation angles $\theta_s$ on the circular cylinder at different Gr given by RTLBFS are compared with the reference results given by Yang et al. [32] and Wang et al. [24]. Here, the average Nusselt number is defined as

$$\overline{\mathrm{Nu}} = \frac{-D}{2\pi(T_h - T_l)} \int_0^{2\pi} \left.\frac{\partial T}{\partial n}\right|_{r=D/2} d\theta \tag{73}$$

As shown in Table 4, the RTLBFS results are in good accordance with the reference data [24, 32].

**Table 4.** Comparison of average Nusselt number and separation angle on a cylinder for mixed convection at Re=20 and various Gr.

| Gr | $\overline{\mathrm{Nu}}$ | | | $\theta_s$ | | |
|---|---|---|---|---|---|---|
| | Wang et al. [24] | Yang et al. [32] | Present | Wang et al. [24] | Yang et al. [32] | Present |
| 0 | 2.523 | 2.454 | 2.438 | 43.19 | 43.59 | 43.65 |
| 100 | 2.640 | 2.655 | 2.653 | 29.73 | 30.01 | 29.91 |
| 800 | 3.208 | 3.201 | 3.202 | 0 | 0 | 0 |

| 1600 | 3.554 | 3.508 | 3.508 | 0 | 0 | 0 |

## 4.4. 3D natural convection in a cubic cavity

To validate the generality of RTLBFs for simulating thermal flows in both 2D and 3D space, 3D natural convection in a cubic cavity is simulated in this subsection. The physical model, i.e., a cubic cavity of size $L\times L\times L$, is shown in Fig. 16. All walls of the cavity are stationary and the no-slip boundary is adopted. The temperatures of the left and right walls are set as $T_l = 0$ and $T_h = 1$, respectively, while the other four walls are adiabatic.

The problem is characterized by the Rayleigh number Ra and the Prandtl number Pr. Their definitions are given in Eq. (63). In the present simulation, four Ra ($10^3$, $10^4$, $10^5$, $10^6$) are chosen, Pr and the characteristic velocity are set as Pr=0.71 and $V_c = 0.1$. For all the cases, the non-uniform mesh of size 80×80×80 is chosen. The coordinates of the mesh nodes are determined by

$$\begin{cases} x_i = \left\{\tanh(\phi) + \tanh\left[2\phi(\frac{i}{M} - 0.5)\right]\right\} \frac{L}{2\tanh(\phi)} & (i = 0,1,\cdots,M) \\ y_j = \left\{\tanh(\phi) + \tanh\left[2\phi(\frac{j}{N} - 0.5)\right]\right\} \frac{L}{2\tanh(\phi)} & (j = 0,1,\cdots,N) \\ z_k = \left\{\tanh(\phi) + \tanh\left[2\phi(\frac{k}{O} - 0.5)\right]\right\} \frac{L}{2\tanh(\phi)} & (k = 0,1,\cdots,O) \end{cases} \quad (74)$$

where $\phi = 1.8$.

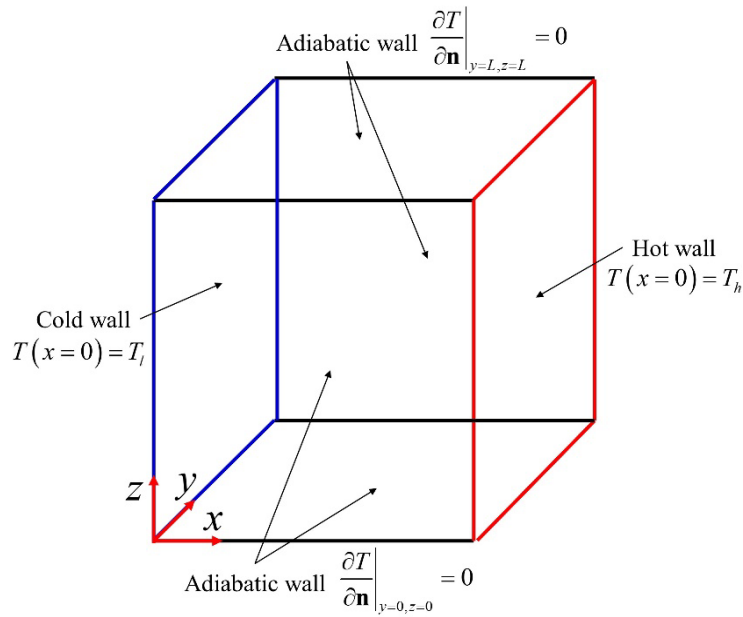

**Figure 16.** Physical model of the 3D natural convection in a cubic cavity.

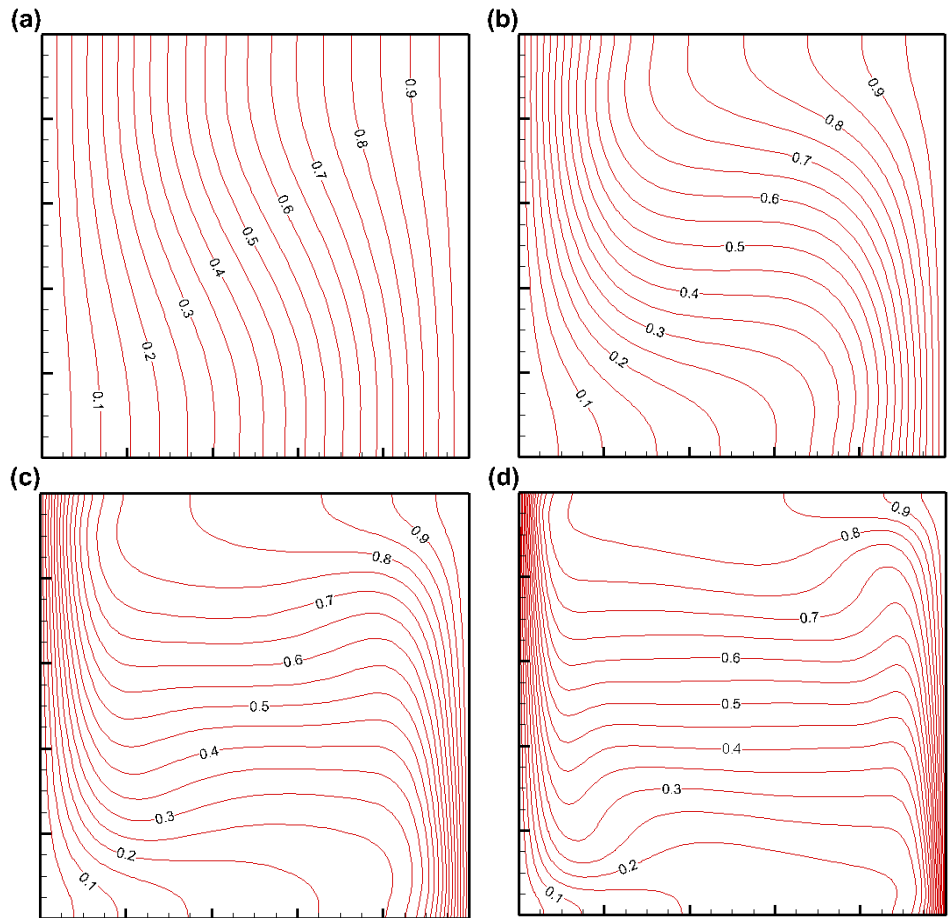

**Figure 17.** Temperature contours on the plane of *y*=0.5*L* for the 3D lid-driven cavity flow at different Ra. (a) Ra=$10^3$, (b) Ra=$10^4$, (b) Ra=$10^5$, (b) Ra=$10^6$.

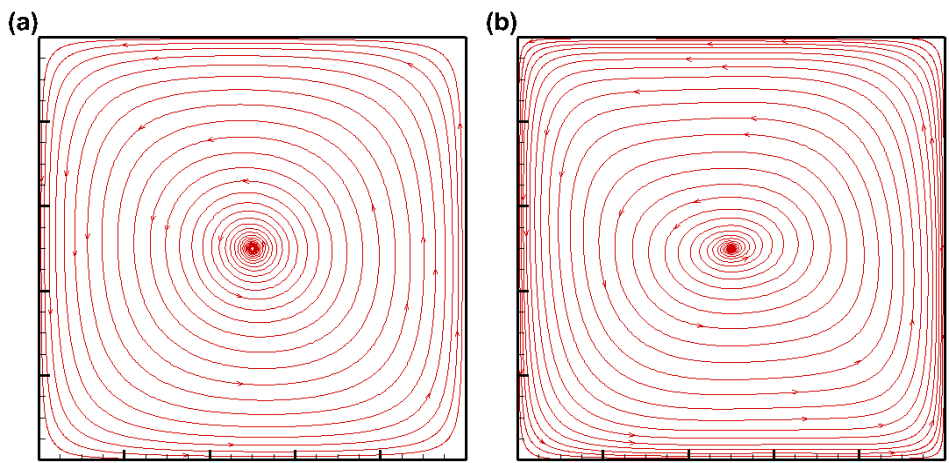

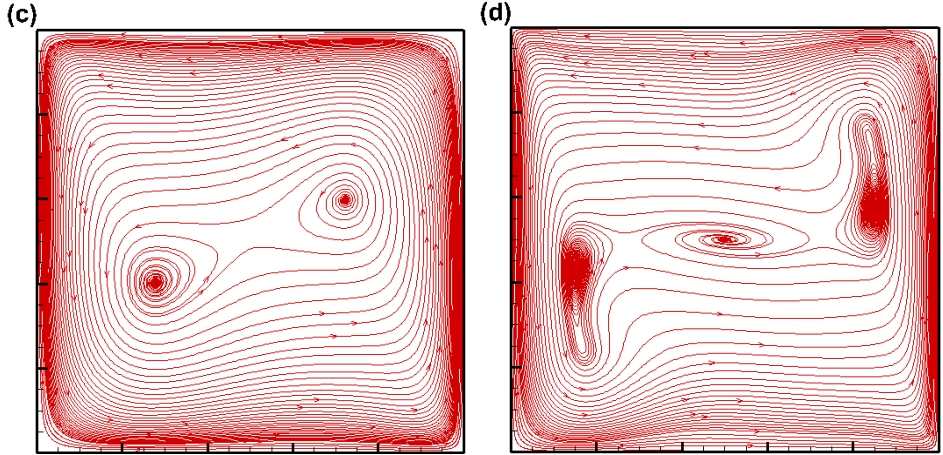

**Figure 18.** Streamlines on the plane of *y*=0.5*L* for the 3D lid-driven cavity flow at different Ra. (a) Ra=$10^3$, (b) Ra=$10^4$, (b) Ra=$10^5$, (b) Ra=$10^6$.

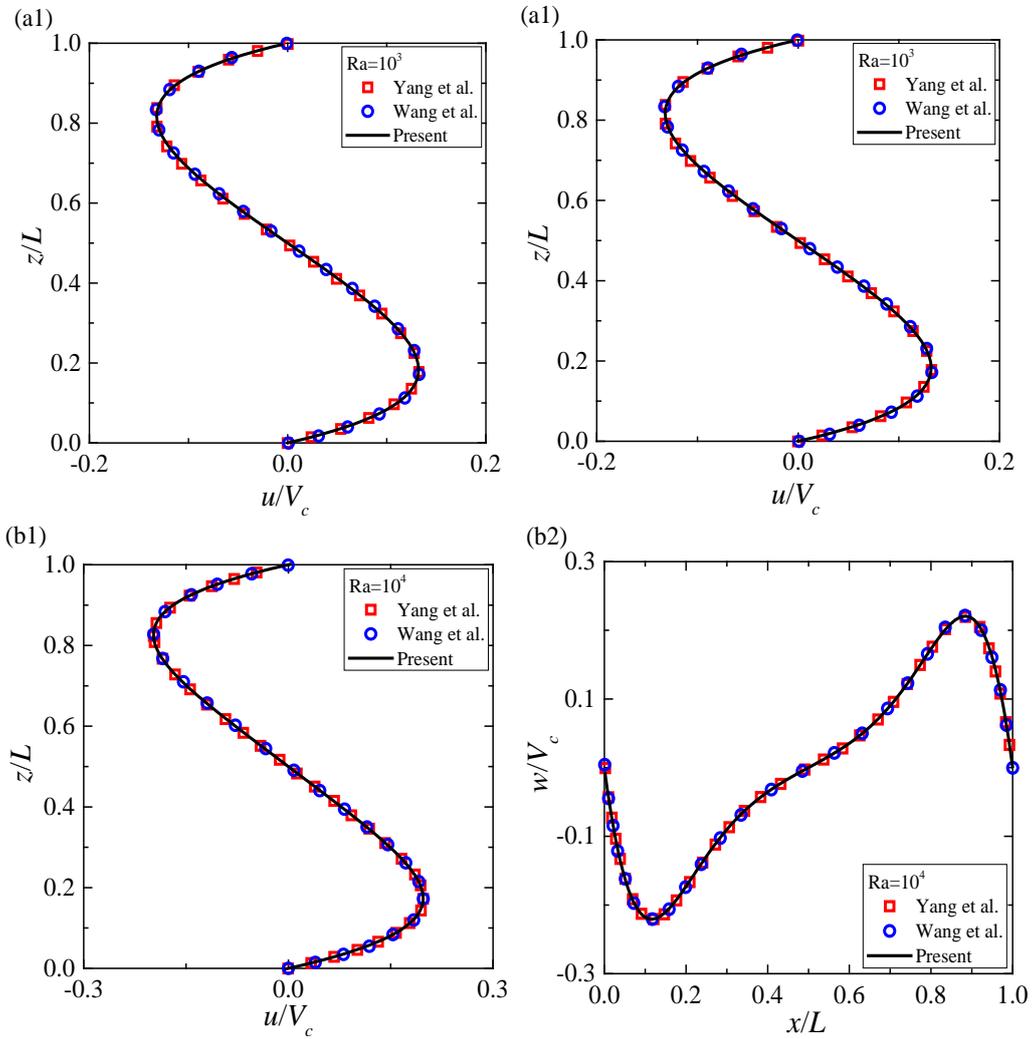

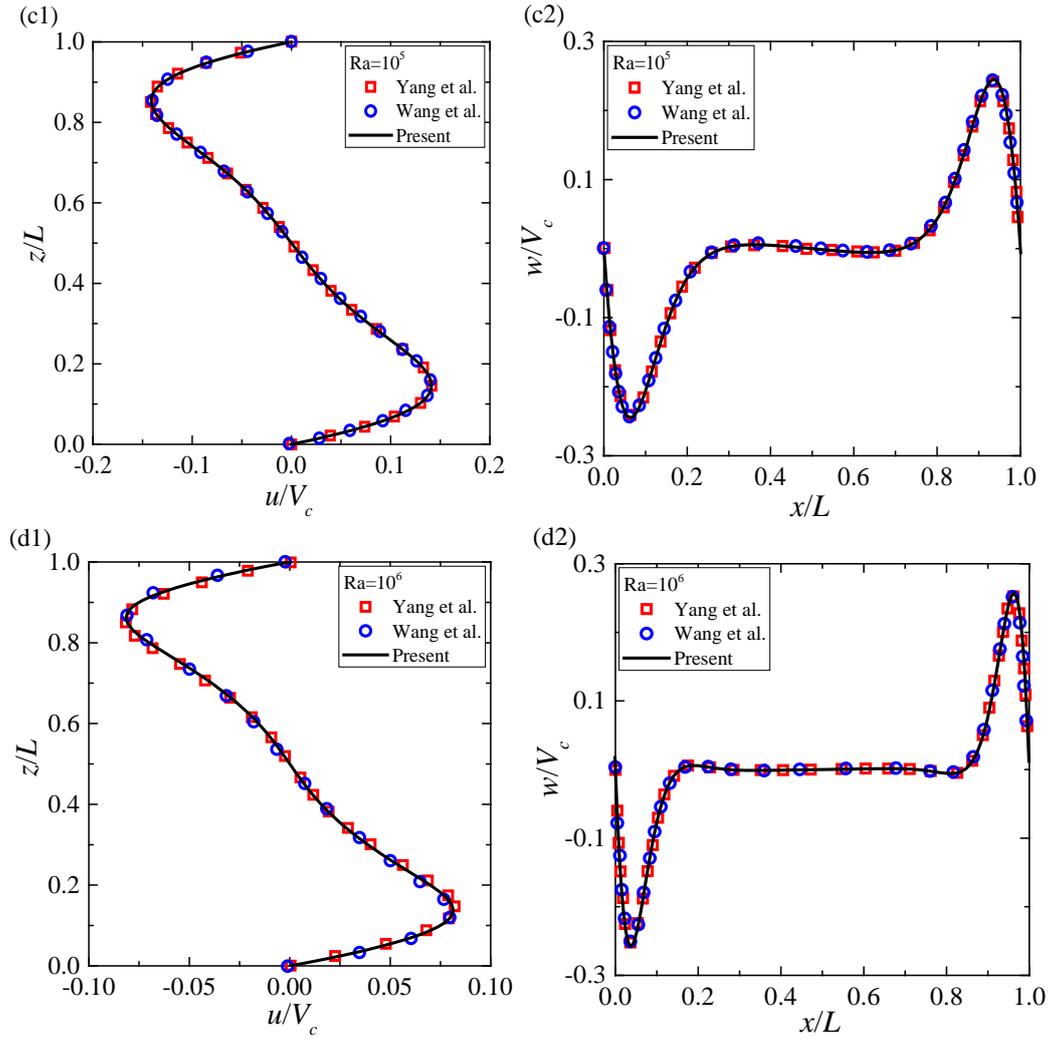

**Figure 19.** The *u* profiles along *x*=0.5*L*, *y*=0.5*L*, and *w* profiles along *y*=0.5*L*, *z*=0.5*L* for the 3D lid-driven cavity flow at different Ra, together with the reference results given by Yang et al. [37] and Wang et al. [25].

The isotherms and streamlines on the symmetric plane *y*=0.5*L* at different Ra are given in Figs. 17 and 18, respectively. With increasing Ra, the isotherms near the left/right boundaries become denser, which indicates larger heat transfer rates in these areas. Besides, the isotherms given by RTLBFS show good agreement with those given by Chen et al. [21] and Yang et al. [37]. The streamlines obtained by RTLBFS also match well with those given by Chen et al. [21] and Yang et al. [37].

To validate the results in quantity, the *u* profiles along *x*=0.5*L*, *y*=0.5*L*, and the *w* profiles along *y*=0.5*L*, *z*=0.5*L* at Ra=$10^7$ and $10^8$ obtained by RTLBFS are compared with the reference results given by Yang et al. [37] and Wang et al. [25]. As shown in Fig. 19, good agreement can be observed between the results given by RTLBFS and the reference results [26, 32].

Moreover, several characteristic parameters, which include the maximum *u* along *x*=0.5*L*, *y*=0.5*L* and the corresponding *z* coordinate, the maximum *w* along *y*=0.5*L*, *z*=0.5*L* and the corresponding *x* coordinate, the mean Nusselt number $Nu_m$ along *x*=*L*, *y*=0.5*L*, and the overall Nusselt number $Nu_o$ on the heated wall, are chosen for comparison. The definitions of $Nu_m$ and $Nu_o$ are respectively given as

$$\mathrm{Nu}_m = \frac{L}{\Delta T} \int_{z=0}^{L} \partial_x T dz, \tag{75}$$

$$\mathrm{Nu}_o = \frac{L}{\Delta T} \int_{y=0}^{L} \int_{z=0}^{L} \partial_x T dy dz. \tag{76}$$

As shown in Table 5, the characteristic parameters obtained by RTLBFS agree well with the reference results [21, 25, 38]. These results validate the ability of RTLBFS for simulating 3D thermal flows.

**Table 5.** Comparison of representative parameters on the $y=0.5L$ plane for natural convection in a cubic cavity.

| Parameters | References | Ra=$10^3$ | Ra=$10^4$ | Ra=$10^5$ | Ra=$10^6$ |
|---|---|---|---|---|---|
| $(u_{x,y=0.5L}L/\chi)_{max}$ | Fusegi et al. [38] | 0.131 | 0.201 | 0.147 | 0.084 |
| | Wang et al. [25] | 0.132 | 0.200 | 0.142 | 0.082 |
| | Chen et al. [21] | 0.132 | 0.198 | 0.140 | 0.077 |
| | Present | 0.132 | 0.198 | 0.141 | 0.0812 |
| $z/L$ | Fusegi et al. [38] | 0.200 | 0.183 | 0.145 | 0.144 |
| | Wang et al. [25] | 0.187 | 0.176 | 0.147 | 0.146 |
| | Chen et al. [21] | 0.188 | 0.175 | 0.150 | 0.138 |
| | Present | 0.185 | 0.175 | 0.147 | 0.138 |
| $(w_{y,z=0.5L}L/\chi)_{max}$ | Fusegi et al. [38] | 0.132 | 0.225 | 0.247 | 0.259 |
| | Wang et al. [25] | 0.133 | 0.221 | 0.244 | 0.253 |
| | Chen et al. [21] | 0.133 | 0.221 | 0.245 | 0.257 |
| | Present | 0.133 | 0.220 | 0.245 | 0.257 |
| $x/L$ | Fusegi et al. [38] | 0.833 | 0.883 | 0.935 | 0.967 |
| | Wang et al. [25] | 0.829 | 0.885 | 0.932 | 0.968 |
| | Chen et al. [21] | 0.825 | 0.883 | 0.933 | 0.964 |
| | Present | 0.825 | 0.885 | 0.936 | 0.963 |
| $\mathrm{Nu}_m$ | Fusegi et al. [38] | 1.105 | 2.302 | 4.646 | 9.012 |
| | Wang et al. [25] | 1.092 | 2.289 | 4.622 | 8.921 |
| | Chen et al. [21] | 1.087 | 2.250 | 4.623 | 8.801 |
| | Present | 1.089 | 2.251 | 4.613 | 8.882 |
| $\mathrm{Nu}_o$ | Fusegi et al. [38] | 1.085 | 2.085 | 4.361 | 8.770 |
| | Wang et al. [25] | 1.071 | 2.062 | 4.344 | 8.684 |
| | Chen et al. [21] | 1.069 | 2.054 | 4.361 | 8.601 |
| | Present | 1.071 | 2.055 | 4.337 | 8.643 |

Besides, to compare the computational efficiency of RTLBFS and TLBFS, the computational time costs $t_{cost}$ for a dimensionless simulation time interval $t^* = V_c t/L = 1$ of the two methods are compared. In the comparison, Ra=$10^6$ is adopted and the computational parameters of the two methods are set the same as above. The series programs with similar structures run on the same computer one by one to obtain the $t_{cost}$ of the two methods. As shown in Table 6, the $t_{cost}$ of RTLBFS is only 2.84% less than that of TLBFS, which indicates that the computational efficiency of the RTLBFS is competitive with TLBFS.

**Table 6.** Comparison of the $t_{cost}$ of RTLBFS and TLBFS for 3D natural convection in a cubic

cavity at Ra=$10^6$. The $t_{cost}$ ratio is defined as the ratio of the present $t_{cost}$ to that of TLBFS.

| Methods | Mesh size | $t_{cost}$ (s) | $t_{cost}$ ratio |
|---|---|---|---|
| TLBFS | 80×80×80 (non-uniform) | 60585.58 | 100% |
| RTLBFS | 80×80×80 (non-uniform) | 58864.12 | 97.16% |

### 4.5. 3D natural convection in a concentric annulus

To show the flexibility of RTLBFS for 3D curved boundary, 3D natural convection in a concentric annulus is simulated. The physical model is presented in Fig. 20. The radii of the inner and outer cylinders are $R_i$ and $R_o$, respectively, and the height of the annulus is $H$. The temperatures on the inner and outer cylinder surfaces are set as $T_i = 1$ and $T_o = 0$, respectively, while the top and bottom walls are adiabatic. For the velocity boundary condition, all walls are stationary and the no-slip boundary condition is adopted. The geometry of the physical model can be determined by two aspect ratios $Ar = R_o/R_i$ and $\eta = H/(R_o - R_i)$. The problem can be characterized by the Prandtl number Pr and the Rayleigh number Ra which are respectively defined as:

$$\Pr = \upsilon/\chi, \quad \mathrm{Ra} = \frac{\beta g (T_i - T_o)(R_o - R_i)^3}{\upsilon \chi} = \frac{V_c^2 (R_o - R_i)^2}{\upsilon \chi}. \tag{77}$$

where $V_c$ is the characteristic velocity defined as $V_c = \sqrt{\beta g (T_i - T_o)(R_o - R_i)}$. In the present subsection, we set $Ar = 2$, $\eta = 2$, $\Pr = 0.71$, and $V_c = 0.1$, and two Ra ($10^4$ and $10^5$) are considered. The mesh size used for simulations is 120×1×240. It is noted that since the physical model is axisymmetric, only 1 mesh layer is needed in the azimuthal direction. By doing this, the computational efficiency can be improved significantly.

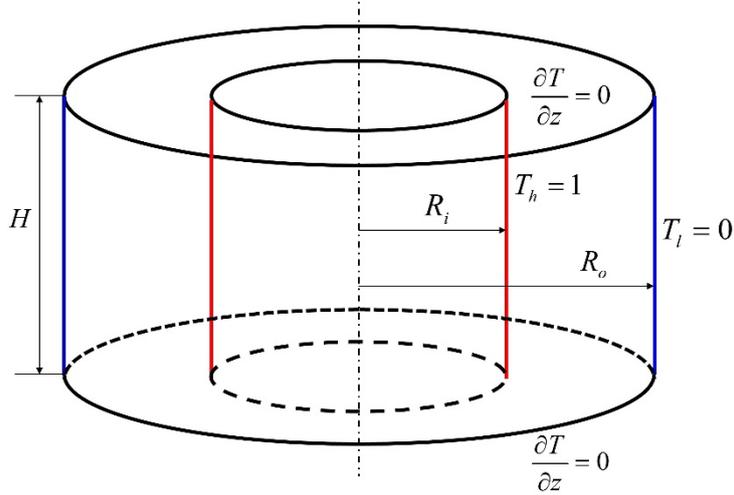

**Figure 20.** Schematic diagram of the 3D natural convection in a concentric annulus.

The steady isotherms at the *r-z* plane of $\mathrm{Ra} = 10^4$ and $10^5$ are depicted in Fig. 21. When Ra increases from $10^4$ to $10^5$, the isotherms near the hot/cool walls become denser, which represents a higher heat transfer rate. Fig. 22 shows the steady streamlines at the *r-z* plane of $\mathrm{Ra} = 10^4$ and $10^5$. For $\mathrm{Ra} = 10^4$, only one vortex appears in the flow field, while for $Ra = 10^5$, three vortices coexist. Besides, both isotherms and streamlines given by RTLBFS agree well with those given by Wang et al. [39] and Chen et al. [21].

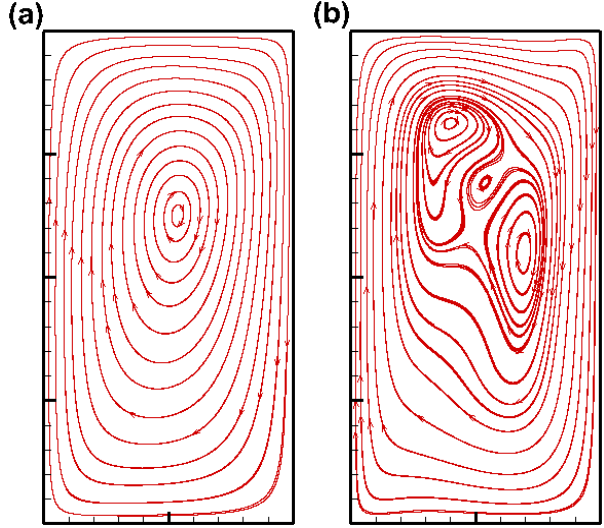

**Figure 21.** Streamlines on the *r-z* plane of the 3D natural convection in a concentric annulus at different Ra. (a) Ra=$10^4$, (b) Ra=$10^5$.

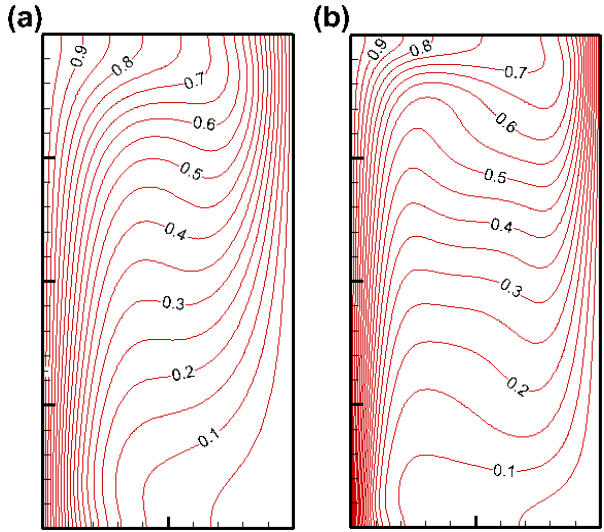

**Figure 22.** Isotherms on the *r-z* plane of the 3D natural convection in a concentric annulus at different Ra. (a) Ra=$10^4$, (b) Ra=$10^5$.

To quantitatively validate RTLBFS for this problem, the average Nusselt numbers of the inner and the outer cylinder surfaces given by RLBFS are compared with the reference data given by Wang et al. [39], Chen et al. [21], and Li et al. [40]. The average Nusselt numbers of the inner and the outer cylinder surfaces are defined as

$$\overline{\mathrm{Nu}}_{i,o} = -\frac{R_{i,o}}{H(T_i - T_o)} \int_0^H \left.\frac{\partial T}{\partial r}\right|_{i,o} dz, \qquad (78)$$

where the subscripts *i* and *o* represent the variables of the inner and outer surfaces. As shown in Table 7, the calculated average Nusselt numbers given by RLBFS agree well with the reference data [21, 39, 40]. The good agreement validates the ability of RTLBFS in tackling the 3D curved boundary.

Table 7. Comparison of average Nusselt numbers for the 3D natural convection in a concentric annulus.

| Ra | Wang et al. [39] | | Chen et al. [21] | | Li et al. [40] | | Present | |
|---|---|---|---|---|---|---|---|---|
| | $\overline{Nu}_i$ | $\overline{Nu}_o$ | $\overline{Nu}_i$ | $\overline{Nu}_o$ | $\overline{Nu}_i$ | $\overline{Nu}_o$ | $\overline{Nu}_i$ | $\overline{Nu}_o$ |
| $10^4$ | 3.220 | 3.217 | 3.228 | 3.193 | 3.126 | 3.216 | 3.212 | 3.222 |
| $10^5$ | 5.815 | 5.808 | 5.731 | 5.712 | 5.798 | 5.798 | 5.791 | 5.810 |

## 5. Conclusions

Based on the thermal lattice Boltzmann model, TLBFS has been proposed to overcome the drawbacks of TLBM including the limitation of uniform mesh, coupled time step with mesh size, and the extra memory size. Since TLBFS only involves macroscopic variables, it belongs to a macroscopic model which simulates incompressible thermal flow with the weakly compressible model. In the perspective of macroscopic scale, directly solving weakly compressible models with the central difference scheme, in general, suffers serious numerical stability. However, TLBFS is proven to have good numerical stability for high Rayleigh number thermal flows. Revealing the mechanism of the good numerical stability of TLBFS for high Rayleigh number thermal flows motivates the present work.

By approximating the actual computational process of TLBFS, MEs-TLBFS with actual numerical dissipation terms are derived firstly. By solving MEs-TLBFS with FVM, RTLBFS is proposed. The comparison between MEs-TLBFS and the standard governing equations indicates that TLBFS includes some additional small terms. Detailed numerical tests prove that these additional small terms are essential to stabilize numerical computation. By retaining these additional small terms, RTLBFS can preserve the good numerical stability of TLBFS for high Rayleigh number flows very well. Therefore, it can be concluded that these additional small terms include some numerical dissipation terms for stabilizing computation, which is the mechanism of the good numerical stability of TLBFS. The result indicates that the good numerical stability of TLBFS can be well explained in the perspective of macroscopic scale, rather than by introducing the mecroscopic theory of thermal lattice Boltzmann models.

More numerical investigations are made to test RTLBFS, and the characteristics of RTLBFS are summarized as follows:

(a) RTLBFS can recover the accuracy of TLBFS well.

(b) RTLBFS is also applicable to simulate both steady and unsteady, 2D and 3D thermal flows, and tackle curved boundaries with non-uniform meshes.

(c) The computational efficiency of the RTLBFS is competitive with TLBFS.

In conclusion, RTLBFS has similar performances as TLBFS for stability, accuracy, flexibility, and efficiency. However, it has a clear mechanism to achieve good numerical stability for high Rayleigh number flows.

## Acknowledgments

The author P Yu would like to thank the financial support from Shenzhen Science and Technology Innovation Commission (Grant No. JCYJ20180504165704491), Guangdong Provincial Key Laboratory of Turbulence Research and Applications (Grant No. 2019B21203001), and the Department of Education of Guangdong Province (Grant No. 2020KZDZX1185). This work is supported by Center for Computational Science and Engineering of Southern University of Science and Technology.

**Nomenclature**

$c_p$ specific heat at constant pressure ($kJ/(kg \cdot K)$)

$c_s$ lattice sound speed ($m/s$)

$D$ diameter ($m$)

$\mathbf{e}_i$ discrete velocity in direction $i$ ($m/s$)

Fo Fourier number $Fo = t\chi/L^2$

$Fo_{\Delta x}$ mesh Fourier number $Fo_{\Delta x} = t\chi/\Delta x^2$

$f_i$ density distribution function ($kg/m^3$)

$f_i^{eq}$ equilibrium density distribution function ($kg/m^3$)

$g$ gravitational acceleration ($m/s^2$)

Gr Rayleigh number $Gr = g\beta(T_h - T_c)D^3/\upsilon^2$

$g_i$ energy distribution function ($J/m^3$)

$g_i^{eq}$ equilibrium energy distribution function ($J/m^3$)

$H$ height ($m$)
$L$ characteristic length ($m$)

$N_0$ grid number

$Nu_a$ average Nusselt number

$\overline{Nu_i}$ average Nusselt number in the inner cylinder surface

$Nu_m$ mean Nusselt number Num along $x = 0.5L$, $y = 0.5L$

$\overline{Nu_o}$ average Nusselt number in the outer cylinder surface

$Nu_o$ overall Nusselt number on the heated wall

$Nu_{1/2}$ average Nusselt number along $x = 0.5L$

$P$ pressure ($kg/(m \cdot s)$)

$P_\alpha$ mass flux ($kg/(m^2 \cdot s)$)

Pe Peclect number $Pe = u_0 L/\chi$

Pr Prandtl number $\text{Pr} = \upsilon/\chi$

$Q_\alpha$ heat flux $i$ ($\text{W}/\text{m}^2$)

$R$ radius (m)

Ra Rayleigh number $\text{Ra} = g\beta(T_h - T_c)H^3/(\upsilon\chi)$

$T$ temperature (K)

$T_h$ high temperature (K)

$T_l$ low temperature (K)

$T^a$ temperature of analytical solution (K)

**u** velocity (m/s)

$u$ velocity component in $x$ direction (m/s)

$u_\infty$ free stream velocity (m/s)

$v$ velocity component in $y$ direction (m/s)

$V_c$ characteristic velocity (m/s)

$w_i$ weight coefficients in direction $i$

$x$ coordinate component (m)
**x** position (m)

$\mathbf{x}_L$ position of the left cell center (m)

$\mathbf{x}_R$ position of the right cell center (m)

$\mathbf{x}_S$ position of the interface center (m)

$y$ coordinate component (m)
$z$ coordinate component (m)

*Greek symbols*

$\beta$ volume expansivity (1/K)

$\chi$ thermal diffusivity ($\text{m}^2/\text{s}$)

$\delta t$ streaming time step (s)
$\varepsilon$ a small parameter
$\theta$ dimensionless temperature

$\rho$ density ($\text{kg}/\text{m}^3$)

$\rho_\infty$ free stream density ($\text{kg}/\text{m}^3$)

$\tau_f$ relaxation time of density distribution function

$\tau_g$ relaxation time of energy distribution function

$\upsilon$ kinematic viscosity ($\text{m}^2/\text{s}$)

$\psi$ streamfunction ($\text{m}^2/\text{s}$)

$\Delta x$ lattice space ($\text{m}$)

$\Delta t$ macroscopic time step ($\text{m}$)

$\Delta V$ control volume ($\text{m}^3$)

$\Delta S$ interface area ($\text{m}^2$)

$\Pi_{\alpha\beta}$ momentum flux ($\text{kg}/(\text{m}\cdot\text{s}^2)$)

*Superscript*
$*$ predicted variables

*Subscripts*
$\alpha$, $\beta$ coordinate component